\documentstyle[11pt,rotate,epsfig]{article}
\begin{document}
\pagenumbering{arabic}
\setcounter{page}{1}
\thispagestyle{empty}
\pagestyle{plain}
\thispagestyle{empty}
\setlength{\baselineskip} {2.5ex}
\def\ie{{\it i.e.} }
\def\etal{{\it et al.}}
\def\sa{$S/A$ }
\def\pt{$p_t$ }
\def\et{$E_t$ }
\def\fm3{fm$^3$}
\def\fmm3{fm$^{-3}$}
\def\Journal#1#2#3#4{{#1} {\bf #2}, #3 (#4)}
\def\NCA{\em Nuovo Cimento}
\def\NIM{\em Nucl. Instr. Meth.}
\def\IJMPA{{\em Int. Jour. Mod. Phys.} A}
\def\NIMA{{\em Nucl. Instr. Meth.} A}
\def\NPA{{\em Nucl. Phys.} A}
\def\NPB{{\em Nucl. Phys.} B}
\def\PLB{{\em Phys. Lett.}  B}
\def\SJPN{\em Sov. Jour. Part. Nucl.}
\def\SJNP{\em Sov. Jour. Nucl. Phys.}
\def\PRL{\em Phys. Rev. Lett.}
\def\PR {\em Phys. Rev.}
\def\PRC{{\em Phys. Rev.} C}
\def\PRD{{\em Phys. Rev.} D}
\def\ZPC{{\em Z. Phys.} C}
\def \ap#1#2#3{{\it Ann. Phys. (N.Y.)} {\bf#1} (#3) #2}
\def \apny#1#2#3{{\it Ann.~Phys.~(N.Y.)} {\bf#1} (#3) #2}
\def \app#1#2#3{{\it Acta Physica Polonica} {\bf#1} (#3) #2}
\def \arnps#1#2#3{{\it Ann. Rev. Nucl. Part. Sci.} {\bf#1} (#3) #2}
\def \arns#1#2#3{{\it Ann. Rev. Nucl. Sci.} {\bf#1} (#3) #2}
\def \cn{Collaboration}
\def \ib{{\it ibid.}~}
\def \ibj#1#2#3{{\it ibid.} {\bf#1} (#3) #2}
\def \ijmpa#1#2#3{{\it Int.~J. Mod.~Phys.}~A {\bf#1} (#3) #2}
\def \ite{{\it et al.}}
\def \jpg#1#2#3{{\it J. Phys.} G {\bf#1} (#3) #2}
\def \kdvs#1#2#3{{\it Kong.~Danske Vid.~Selsk., Matt-fys.~Medd.} {\bf #1}
(#3) No #2}
\def \mpla #1#2#3{{\it Mod. Phys. Lett.} A {\bf#1} (#3) #2}
\def \nc#1#2#3{{\it Nuovo Cim.} {\bf#1} (#3) #2}
\def \np#1#2#3{{\it Nucl. Phys.} {\bf#1} (#3) #2}
\def \pisma#1#2#3#4{{\it Pis'ma Zh. Eksp. Teor. Fiz.} {\bf#1} (#3) #2 [{\it
JETP Lett.} {\bf#1} (#3) #4]}
\def \pl#1#2#3{{\it Phys. Lett.} {\bf#1} (#3) #2}
\def \plb#1#2#3{{\it Phys. Lett.} B {\bf#1} (#3) #2}
\def \ppnp#1#2#3{{\it Prog. Part. Nucl. Phys.} {\bf#1} (#3) #2}
\def \pr#1#2#3{{\it Phys. Rev.} {\bf#1} (#3) #2}
\def \prd#1#2#3{{\it Phys. Rev.} D {\bf#1} (#3) #2}
\def \prl#1#2#3{{\it Phys. Rev. Lett.} {\bf#1} (#3) #2}
\def \prp#1#2#3{{\it Phys. Rep.} {\bf#1} (#3) #2}
\def \ptp#1#2#3{{\it Prog. Theor. Phys.} {\bf#1} (#3) #2}
\def \rmp#1#2#3{{\it Rev. Mod. Phys.} {\bf#1} (#3) #2}
\def \rp#1{~~~~~\ldots\ldots{\rm rp~}{#1}~~~~~}
\def \yaf#1#2#3#4{{\it Yad. Fiz.} {\bf#1} (#3) #2 [Sov. J. Nucl. Phys. {\bf #1}
 (#3) #4]}
\def \zhetf#1#2#3#4#5#6{{\it Zh. Eksp. Teor. Fiz.} {\bf #1} (#3) #2 [Sov.
Phys. - JETP {\bf #4} (#6) #5]}
\def \zhetfl#1#2#3#4{{\it Pis'ma Zh. Eksp. Teor. Fiz.} {\bf #1} (#3) #2 [JETP
Letters {\bf #1} (#3) #4]}
\def \zp#1#2#3{{\it Zeit. Phys.} {\bf#1} (#3) #2}
\def \zpc#1#2#3{{\it Zeit. Phys.} C {\bf#1} (#3) #2}
\newcommand{\asmu}{\alpha_s(\mu^2)}
\newcommand \beq{\begin{eqnarray}}
\newcommand \eeq{\end{eqnarray}}
\newcommand \ga{\raisebox{-.5ex}{$\stackrel{>}{\sim}$}}
\newcommand \la{\raisebox{-.5ex}{$\stackrel{<}{\sim}$}}
\newcommand \eps{\varepsilon}
\newcommand \ep{\varepsilon_{\bf p}}
\newcommand \Del{\bigtriangledown}
\newcommand \ps {P(\sigma)}
\newcommand \s{\bar{\sigma}}
\newcommand \rhot{\tilde\rho}
\newcommand \al{\tilde\alpha}
\newcommand \be{\tilde\beta}
\newcommand \sj{\sigma_{pj}}
\newcommand \sjj{\sigma_{pj'}}
\newcommand \sij{\sigma_{ij}}
\newcommand \os{\omega_\sigma}
\newcommand{\ve}[1]{\bf {#1}}
\newcommand{\ave}[1]{{\bf <}\!\!{#1}\!\!{\bf >}}
\newcommand{\av}[1]{\langle{#1}\rangle}
\newcommand{\rep}[1]{(\ref{#1})}
\newcommand{\bea}{\begin{eqnarray}}
\newcommand{\eea}{\end{eqnarray}}
\newcommand{\ket}[1]{|{#1}\rangle}
\newcommand{\bra}[1]{\langle{#1}|}
\newcommand{\tti}{\tilde{t}}
\newcommand{\half}{{\scriptstyle \frac12 }}
\newcommand{\br}{{\bf r}}
\newcommand{\ai}{\alpha_i}
\newcommand{\aj}{\alpha_j}
\newcommand{\am}{\alpha_m}
\newcommand{\dsdtdm}{d\sigma/dt dM^2}
\newcommand \sd{\sigma_{diff}}
\newcommand{\ks}{\kappa_\sigma}
\newcommand{\rdeg}{$^\circ$}

\vspace{3.cm}

\begin{center}
{Contribution to the Proceedings of the 
Charles U./JINR and International U. (Dubna) 
CERN COMPASS Summer School,\\ 
Charles University, Prague, Czech Republic, August 1997,\\
Eds. M. Chavleishvili and M. Finger\\
Tel Aviv U. Preprint TAUP-2473-98\\
HEP-EX Archive http://xxx.lanl.gov, HEP-EX/9801011}

\vspace{0.6in}

{\Large\bf
Primakoff Physics for CERN COMPASS Hadron Beam:\\
Hadron Polarizabilities, Hybrid Mesons, Chiral Anomaly, Meson
Radiative Transitions}

\vspace{0.3in}
{Murray A. Moinester, Victor Steiner, \\
School of Physics and Astronomy,\\ R. and B. Sackler Faculty 
of Exact Sciences,\\ Tel Aviv University, 69978 Ramat Aviv, Israel\\
e-mail: murraym$@$silly.tau.ac.il, steiner$@$gluon.tau.ac.il} 

\vspace{1cm}

{\large\bf  Abstract}
\end{center}
\small

We describe a hadron physics program attainable with a partially instrumented CERN
COMPASS spectrometer, involving tracking detectors and  moderate-size ECAL2/HCAL2
calorimeters. COMPASS can realize a state-of-the-art hadron beam physics program
based on hadron polarizability, hybrid mesons, chiral anomaly, and meson radiative
transition studies. We review here the physics motivation for this hadron beam
program. We describe the beam, detector, trigger requirements, and hardware/software
requirements for this program. The triggers for all this physics can be implemented
for simultaneous data taking. The program is based on using a hadron beam
(positive/negative pion, kaon, proton) in COMPASS.

\normalsize

\newpage
\section{Physics Review}

The approved COMPASS physics program \cite {compass} includes studies of
$\gamma$-hadron Primakoff interactions using 50-280 GeV/c negative beams (pions,
kaons) and positive beams (pions, kaons, protons) together with a virtual photon
target in dedicated data runs. Pion and kaon and proton polarizabilities, hybrid
mesons, the chiral anomaly, and radiative transitions can be studied in this way, and
can provide significant tests of QCD and chiral perturbation theory ($\chi$PT)
predictions. All of these subjects may be studied at the same time. Some COMPASS
studies given in this report also appear in Refs. \cite {cd2,hadron1}.

\subsection{Hadron Polarizibilities}

For the $\gamma$-$\pi$ interaction at low energy, chiral perturbation theory
($\chi$PT) provides a rigorous way to make predictions; because it stems directly
from QCD and relies only on the solid assumptions of spontaneously broken SU(3)$_L$
$\times$ SU(3)$_R$ chiral symmetry, Lorentz invariance and low momentum transfer.
Unitarity is achieved by adding pion loop corrections to lowest order, and the
resulting infinite divergences are absorbed into physical (renormalized) coupling
constants L$^r_i$ (tree-level coefficients in L$^{(4)}$, see Refs.
\cite{donn1,gass1}). With a perturbative expansion of the effective Lagrangian
limited to terms quartic in the momenta and quark masses (O(p$^4$)), the method
establishes relationships between different processes in terms of the L$^r_i$. For
example, the radiative pion beta decay and electric pion polarizability are expressed
as \cite{donn1}:
\begin{equation}
h_A/h_V = 32\pi^2(L^r_9+L^r_{10}); \bar{\alpha}_{\pi} =
\frac{4\alpha_f}{m_{\pi}F^{2}_{\pi}}(L^r_9+L^r_{10});
\label{eq:L9}
\end{equation}
\noindent
where F$_\pi$ = 93.1  MeV \cite{part}  is the pion decay constant, h$_A$ and h$_V$ 
are the axial vector and vector coupling constants in the decay, and $\alpha_f$ is
the fine structure constant. The experimental ratio \cite{part} h$_A$/h$_V$  = 0.45
$\pm$ 0.06, leads to $\bar{\alpha}_{\pi}$ = -$\bar{\beta}_{\pi}$ = 2.7 $\pm$ 0.4,
where the error shown is due to the uncertainty in the h$_A$/h$_V$ measurement
\cite{babu2}. {\bf All polarizabilities in this paper are expressed in Gaussian units
of 10$^{-43}$ cm$^3$.}

Holstein \cite {hols} showed that meson exchange via a pole diagram involving the
a$_1$(1260) resonance provides the main contribution ($\bar{\alpha}_{\pi}$ = 2.6) to
the polarizability. COMPASS can obtain new high statistics data for radiative
transitions leading from the pion to the a$_1$(1260), and to other meson resonances.

In fact, the a$_1(1260)$ width and the pion polarizability are related to an
interesting question, which is whether or not one can expect gamma ray rates from the
quark gluon plasma to be higher than from the hot hadronic gas phase in relativistic
heavy ion collisions. Xiong, Shuryak, Brown (XSB) calculate photon production from a
hot hadronic gas via the reaction $\pi^- + \rho^0 \rightarrow \pi^- + \gamma$. They
assume that this reaction proceeds through the a$_1$(1260). For a$_1$(1260)
$\rightarrow \pi \gamma$, the experimental width \cite {ziel} is $\Gamma = 0.64 \pm
0.25$ MeV. Xiong, Shuryak, and Brown (XSB) \cite {xsb} estimate this  radiative width
to be $\Gamma$ = 1.4 MeV, more than two times higher than the experimental value
\cite {ziel}. With this estimated width, they calculate the pion polarizability to be
$\bar{\alpha}_{\pi}$ = 1.8. COMPASS can experimentally check the a$_1$ dominance
assumption of XSB, and the consistency of the expected relationship of this radiative
width and the pion polarizability \cite {qgp}.

For the kaon, $\chi$PT predicts \cite {cd2,hols,moin2,cd1,mich} $\bar{\alpha}_{K^-}$
= 0.5 . The kaon polarizability measurements at COMPASS should complement those for
pion polarizabilities for chiral symmetry tests away from the chiral limit. A more
extensive study of kaon polarizabilities was given recently by Ebert and Volkov \cite
{ev}. Until now, only an upper limit \cite {gb} at 90\% confidence was measured (via
energy shifts in heavy Z kaonic atoms) for the $K^-$, with $\bar{\alpha}_{K} \leq
200.$

\begin{figure}[tbc]
\centerline{\epsfig{file=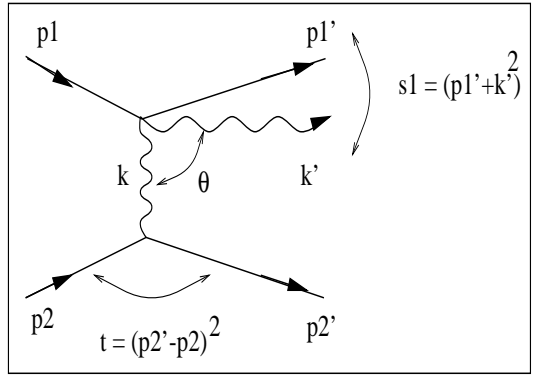,width=7cm,height=5cm}}
\caption{The Primakoff $\gamma$-hadron Compton process and kinematic variables
(4-momenta): p1, p1$^\prime$ = for initial/final hadron, p2, p2$^\prime$ = for 
initial/final target, k, k$^\prime$ = for initial/final gamma, and $\theta$ 
the scattering angle of the $\gamma$ in the alab frame.}
\label{fig:diagram}
\end{figure}

The polarizabilities can be obtained from precise measurements of the $\gamma$-hadron
Compton scattering differential cross sections. Antipov et al. \cite {anti} measured
the $\gamma\pi$ scattering with 40 GeV pions via radiative pion scattering
(Bremsstrahlung) in the nuclear Coulomb field
\begin{equation} \pi^-  +  Z \rightarrow  {\pi^-}'  + Z'  +
\gamma \label{eq:polariz}
\end{equation}
\noindent
where Z is the nuclear charge. Fig.~\ref{fig:diagram} defines the kinematic variables
for such an experiment. The four-momentum of each particle in Eq.~\ref{eq:polariz} is
p1, p2, p1$^\prime$, p2$^\prime$, k$^\prime$, respectively. In the one-photon
exchange domain, this reaction is equivalent to $\gamma + \pi  \rightarrow 
\gamma^\prime + \pi^\prime$, and the four-momentum of the incident virtual photon is
k = p2$-$p2$^\prime$. We have therefore  t = k$^2$ with t the square of the
four-momentum transfer to the nucleus, F(t) the nuclear form factor (essentially
unity at small t, $\sqrt{s}$ the mass of the $\gamma\pi$ final state. The t is
larger than t$_0$, the minimum value of t to produce a mass $\sqrt{s}$, see 
Section~\ref{sec:evt_gen} for details. The momentum modulus $|\vec{k}|$
(essentially equal to p$_T$) of the virtual photon is in the transverse direction,
and is equal and opposite to the momentum p$_T$ transferred to the target nucleus.
The final state $\gamma$ and pion were detected in coincidence. The data selection
criteria required one photon and one charged particle in the final state, their total
four momenta consistent with that of the beam, small four momentum transfer t
to the target nucleus, software cuts on the invariant energy of the final state
$\pi\gamma$ system that are equivalent to choosing effective $\gamma$ energies of
100$-$600 MeV in the pion rest frame (designated hereafter as anti-lab frame or
"alab"), and other position, angle, and energy/momentum conditions.

The virtual photon with four-momentum k=$\{\omega,\vec{k}\}$ has virtual mass
M$^2$=k$^2$=t=$\omega^2-|\vec{k}|^2$. Since t=2M$_Z$[M$_Z$ - E($Z'$,lab)] $<$ 0, 
the virtual
photon mass is imaginary. To approximate real pion Compton scattering, the virtual
photon may be taken to be almost real. The COMPASS trigger and projected statistics
will provide data over effective $\gamma$ energies of 100$-$2000 MeV in the alab
frame.  Such an energy range compared to Antipov et al. allows significantly
increased sensitivity to the polarizabilities. Good resolution in t is important,
since the characteristic signature for Primakoff scattering is low-t, while the
scattering through other processes such as meson and Pomeron exchange have larger t.
Antipov et al. \cite {anti} measured the pion electric polarizability
$\bar{\alpha_{\pi}}$ with low statistics ($\sim$ 7000 events) and found
$\bar{\alpha}_{\pi} = 6.8 \pm 1.4 (\rm{stat}) \pm 1.2 (\rm{syst})$~\cite {anti}. This
result included the assumption that $\bar{\alpha}_{\pi}+\bar{\beta}_{\pi}\approx0.4$,
based on dispersion sum rules~\cite{Pennington}. The value 6.8 reported is far from
the $\chi$PT prediction. More precise measurements of experimental polarizabilities
are needed in order to subject the chiral perturbation techniques of QCD to new and
serious tests.

\subsection{Hybrid Mesons}

The hybrid ($q\bar{q}g$) mesons, along with glueballs ($gg$) are one of the most
amazing consequences of the non-abelian nature of QCD. Detection of these exotic
states is a long-standing experimental puzzle. The most popular approach for the
hybrids search is to look for the "oddballs" - mesons with the quantum numbers not
allowed for the $q\bar{q}$ states, for example J$^{PC}= 1^{-+}$, decaying to $\eta
\pi$, $\eta ' \pi$, $f_1(1285) \pi$, $b_1(1235) \pi$, etc.

From more than a decade of experimental efforts at  IHEP \cite {ihep1,ihep2,ves}, CERN
\cite {na12}, KEK \cite{kek} and BNL \cite {E852}, several hybrid candidates have
been identified. The most recent information came from BNL E852 experiment \cite
{E852} which studied $\pi^- p$ interaction at 18 GeV/c. They reported J$^{PC}=
1^{-+}$ resonant signals in $\eta \pi^-$ and $\eta \pi^0$ systems as well as in
$\pi^+ \pi^- \pi^-$, $\pi^- \pi^0 \pi^0$, $\eta ' \pi^-$ and $f_1(1285) \pi^-$. At
the same time, a VES group \cite {ves} has published analysis of $\eta \pi^-$, $\eta
' \pi^-$, $f_1(1285) \pi^-$, $b_1(1235) \pi^-$ and $\rho \pi^-$ systems production in
$\pi^- Be$ interaction at 37 GeV/c. Although the J$^{PC}= 1^{-+}$ wave is clearly
seen by VES in all channels, there is no indication for the presence of narrow
($\Gamma \sim 0.2~ GeV$) resonance in any of them. But an observed abnormally high
ratio of $\eta ' \pi$ to $\eta \pi$ P-wave is considered as an evidence on hybrid
nature of this exotic wave.

It should be mentioned that the partial wave analysis (PWA) of systems such as $\eta
\pi$ or $\eta ' \pi$ in the mass region below 2 GeV is particularly difficult. This
is so because (1) this region is dominated by the strong $2^+$ "background" (a1
resonance), and (2) that the PWA may give ambiguous results \cite {ihep2} for the
weaker $1^{-+}$ wave. The problem is that the PWA of the $\eta\pi$ system must take
into account S, P and D waves, and the number of observables is not sufficient to
solve unambiguously all equations. Looking at the partial wave solutions as a
function of mass, each partial wave can have as many as eight different curves to
describe its strength and phase, as discussed in ref. \cite {ihep2}. It is therefore
extremely important to have extra information from different hybrid production
mechanisms where the physics is different and such ambiguities may look different.
Only by comparing results of different experiments in this way,  can we establish
unambiguously the existence or non-existence of hybrid (or exotic) meson states.

COMPASS can contribute significantly to the further investigation of hybrids by
studying Primakoff production of  J$^{PC}= 1^{-+}$ $\rhot$ hybrids. The possibilities
for Primakoff production of the $\rhot$ with energetic pion beams, and detection via
different decay channels has been discussed by Zielinski et al. \cite {zihy}, and
Monte Carlo simulations for this physics were carried out for the 600 GeV FNAL SELEX
run \cite {zihy}. Considering vector dominance models, if the $\rhot$ has a 1-10 MeV
branching width into the $\pi\rho$ channel, a branching width of $\rhot$ into the
$\pi\gamma$ channel should be 3-30 keV \cite {zihy}. A hybrid state with such a large
radiative width would be produced at detectable levels through the Primakoff
mechanism in COMPASS. A $\gamma-\pi$ COMPASS trigger should allow observation of the
$\rhot$ via the $\eta\pi^-$ decay mode. With a relative P wave (L=1), the $\eta\pi^-$
system has J$^{PC}= 1^{-+}$. The other decay channels of $\rhot$ may be studied
simultaneously in COMPASS by a relatively simple particle multiplicity trigger (say,
three charged particles in final state).

The evidence presented for the hybrid (pionic) meson offers COMPASS an exceptional
opportunity to take the next steps in this exciting field. COMPASS can study hybrid
meson candidates near 1.4 GeV produced by the Primakoff process. COMPASS should also
be sensitive to pionic hybrids at higher excitation, and also to kaonic hybrids,
which have not yet been reported. We may obtain superior statistics for a hybrid
state if it exists, and via a different production mechanism without possible
complication by hadronic final state interactions. We may also get important data on
the different decay modes for this state. The observation of this hybrid in different
decay modes and in a different experiment would constitute the next important step
following the evidence so far reported.

COMPASS can provide a unique opportunity to investigate QCD exotics, glueballs and
hybrids,  produced via different production mechanisms: central production for
glueballs and Primakoff production for hybrids. Taking into account the very high
beam intensity, fast data acquisition, high acceptance and good resolution of the
COMPASS setup, one can expect from COMPASS the highest statistics and a
"systematics-free" data sample that includes many tests to control possible
systematic errors. The COMPASS effort should significantly improve our understanding
of hybrid and glueball physics.

\subsection{Chiral Axial Anomaly}
 
The Chiral Axial Anomaly can also be studied with 50-280 GeV pion beams with the
same $\pi\gamma$ trigger as used above. For the $\gamma$-$\pi$ interaction, the
O(p$^4$) chiral lagrangian \cite {donn1,gass1} includes Wess-Zumino-Witten (WZW)
terms \cite{wzw,bij3}, which lead to a chiral anomaly term \cite{anti2,wzw,bij3} in
the divergence equations of the currents. This leads directly to interesting
predictions \cite{bij3} for the processes $\pi^0 \rightarrow  2 \gamma$ and $\gamma
\rightarrow  3 \pi$; and other processes as well \cite{bij3}. The two processes
listed are described by the amplitudes F$_{\pi}$ and F$_{3\pi}$, respectively.

The chiral anomaly term leads to a prediction for F$_{\pi}$ and F$_{3\pi}$ in terms
of $N_c$, the number of colors in QCD; and f, the charged pion decay constant. The
O(p$^4$) F$_{\pi}$ prediction is in agreement with experiment \cite {bij3}. The
F$_{3\pi}$ prediction is \cite {ca,hols2}:

\begin{eqnarray}
F_{3\pi} = {N_c (4 \pi \alpha)^{1 \over 2} 
\over 12 \pi^2 f^3} \sim 9.7 \pm 0.2 ~\rm{GeV}^{-3},~ O(p^4). 
\label{eq:f3pi}
\end{eqnarray}
\noindent
The experimental confirmation of this equation would demonstrate that the
O(p$^4$) terms are sufficient to describe F$_{3\pi}$.

The amplitude F$_{3\pi}$ was measured by Antipov et al. \cite{anti2} at
Serpukhov with 40 GeV pions. Their study involved pion production by a pion in
the nuclear Coulomb field via the Primakoff reaction:

\begin{equation}
 \pi^- + Z \rightarrow {\pi^-}' + \pi^0 + Z'.
\label{eq:anomaly}
\end{equation}

In the one-photon exchange domain, Eq. 4 is equivalent to: 

\begin{equation}
 {\pi^-} + \gamma  \rightarrow  {\pi^-}' + {\pi^0},          
\label{eq:eqanomaly}
\end{equation}
\noindent
and the 4-momentum of the virtual photon is k = p$_Z$-p$_{Z'}$. The cross section
formula for the Primakoff reaction depends on $F_{3\pi}^2$. The Antipov et al. data
sample (roughly 200 events) covered the ranges $-$t $<$ 2. $\times 10^{-3}$ (GeV/c)$^2$ and
s$(\pi^-\pi^0) < 10.~m_{\pi}^2$. The small t-range selects events predominantly
associated with the exchange of a virtual photon, for which the target nucleus acts
as a spectator. Diffractive production of the two-pion final state is blocked by
G-parity conservation. The experiment \cite{anti2} yielded 
F$_{3\pi}$=12.9 $\pm$ 0.9 (stat) $\pm$ 0.5 (sys) GeV$^{-3}$. 
This result differs from the O(p$^4$) expectation by
at least two standard deviations; so that the chiral anomaly prediction  at O(p$^4$)
is not confirmed by the available $\gamma \rightarrow 3\pi$ data.
                                 
Bijnens et al. \cite{bij3} studied higher order $\chi$PT corrections in
the abnormal intrinsic parity (anomalous) sector. They included one-loop
diagrams involving one vertex from the WZW term, and tree diagrams from the
O(p$^6$) lagrangian. They determine parameters of the lagrangian via vector
meson dominance (VMD) calculations. The higher order corrections are small for
F$_{\pi}$. For F$_{3\pi}$, they increase the lowest order value from 7\% to
12\%. The one-loop and  O(p$^6$) corrections to F$_{3\pi}$ are comparable in
strength. The loop corrections to F$_{3\pi}$ are not constant over the whole
phase space, due to dependences on the momenta of the 3 pions. The average
effect is roughly 10\%, which then increases the theoretical prediction by 1
GeV$^{-3}$. The prediction is then $F_{3\pi} \sim  10.7$, closer to the 
data. The limited accuracy of the existing data, together with the new
calculations of Bijnens et al., motivate an improved and more precise
experiment.
      
\subsection{Meson Radiative Transitions}

With the same trigger as needed for the above studies, we can  also obtain new
high statistics data for the radiative transitions of incident mesons to
higher excited states; such as from the pion to the $\rho^-$ and from the
K$^{-}$ to the K$^{*-}$. These radiative transition widths are predicted by
vector dominance and quark models. For $\rho \rightarrow \pi \gamma$, the widths
obtained previously \cite{jens,hust,capr} range from 60 keV to 81 keV. For K$^*
\rightarrow K \gamma$, the widths obtained previously are 48 $\pm$ 11 keV \cite
{berg} and 51 $\pm$ 5 keV \cite {chan}. Independent data for these and higher
resonances would be valuable to get higher precision measurements to allow a more
meaningful comparison with theoretical predictions.

With a particle multiplicity trigger, we will also obtain new high statistics data
for radiative transitions leading from the pion to the a$_1$(1260), and to the
a$_2$(1320), and to other resonances or exotics. These radiative transition widths
were studied in the past by different groups by vector dominance and quark models,
but independent data would still be of value. For a$_1$(1260) $\rightarrow \pi
\gamma$, the width given \cite{ziel} is $\Gamma = 0.64 \pm 0.25$ MeV, and for
a$_2$(1320) $\rightarrow \pi \gamma$, the width given \cite{ciha} is $\Gamma = 0.30
\pm$ 0.06 MeV.

\section{Experimental Requirements}

\begin{figure}[tpc]
\centerline{\epsfig{file=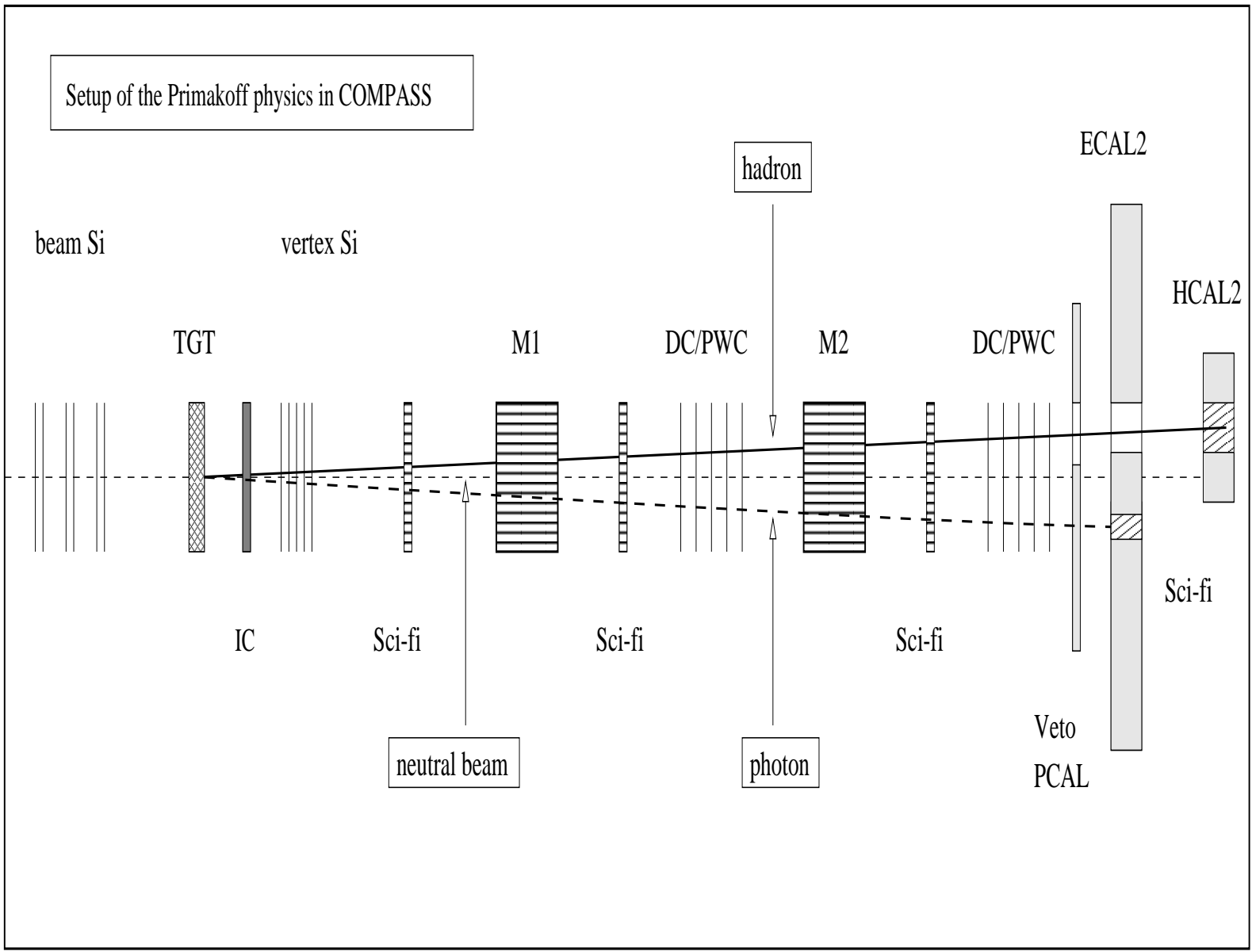,width=16cm,height=21cm}}
\caption{COMPASS setup for polarizability physics.}
\label{fig:setup}
\end{figure}

We consider the beam, detector, trigger and financial requirements for
polarizability, hybrid, and anomaly studies, beginning for illustration with
pion polarizability measurements with a 300 GeV pion beam. Although our illustrative
simulations are given at 300 GeV, the data run will actually be at the maximum
convenient energy for the COMPASS beam line, closer to 280 GeV. The beam energy is
chosen to be maximal, since that pushes the energy spectrum of final state
$\gamma$'s, $\pi^0$'s, and $\eta$'s to be highest, and thereby the detection
acceptance for $\eta$'s for a given size ECAL2 electromagnetic calorimeter will be
maximal.

The reaction $\pi^{-}$ + Z $\rightarrow$  $\pi^{-'}$  + $\gamma$ + Z$'$ is considered
for illustration. Simulations of the other reactions in the COMPASS apparatus are in
progress. The experimental setup is shown schematically in Fig.~\ref{fig:setup}. In
Fig.~\ref{fig:kinematics}, we show the kinematics from our Monte Carlo \cite{pol}
study of the measurement accuracy. Fig.~\ref{fig:acceptance} shows the acceptance for
angular distribution measurements, due to various trigger conditions. In
Fig.~\ref{fig:pi_gamma_correl} we show various correlations between the $\pi$ and
$\gamma$ kinematic variables in the lab frame. In Fig.~\ref{fig:cth_correl} we show
the important correlation between the scattering angle of the $\gamma$ in alab frame
(used to extract the polarizability) and the $\pi$ and $\gamma$ kinematic variables
in the lab frame.

\subsection{Monte Carlo Simulations}

We have carried out Monte Carlo simulations with two codes, POLARIS , an event
generator for polarizability studies and ANOMALY, developed for chiral anomaly
studies\cite{pol}. In this report, we show only the POLARIS results. For hybrid
mesons, simulations \cite {zihy} were carried out for the SELEX apparatus, and need
to be done now for COMPASS.

POLARIS produces events of type:
\begin{equation}
h + Z \rightarrow h' + \gamma + Z'
\end{equation}
\noindent
where $h$ is a pion or kaon, based on the theoretical Primakoff $\gamma$-meson
Compton scattering cross section. The meson and $\gamma$ laboratory variables may be
given gaussian spreads to simulate instrumental errors and acceptance cuts may be
used (optional). Finally, the event is reconstructed from these "measured" values.
The meson polarizability is extracted via a fit of the theoretical cross section to
the scattered photon angular distribution in the projectile (alab) frame. The total
cross section is computed by integrating numerically the differential cross section
$\sigma$(s,t,$\theta$) of the Compton process (see Fig.~\ref{fig:diagram}). The various
techniques used are commented in the code. The relevant files are described in
Table~\ref{tab:polaris}. 

The code ANOMALY produces events of type:
\begin{equation}
h + Z \rightarrow h' + \pi^o + Z'
\end{equation}
\noindent
where $\pi^o$ decays to $2\gamma$, using similar techniques as POLARIS.

\begin{table}
\begin{center}
\begin{tabular}[tbc]{|l|l|}
\hline
File             & Function\\
\hline
polaris.f        & source code\\
pol\_pi\_pb.inp  & input data\\
polaris.dat      & output event file in Geant-like format\\
events.dat       & output binary event file for fast reruning\\
polaris.hbk      & output histogram file\\
mkpol            & to make the executable code\\
gopol            & to run the code\\
\hline
\end{tabular}
\end{center}
\caption{Description of input/output files for the Monte Carlo event generator
POLARIS.}
\label{tab:polaris}
\end{table}

\subsection{Primakoff $\gamma-\pi$ Compton Event Generator\label{sec:evt_gen}}

We give more details regarding the event generator for the radiative scattering of
the pion (pion Bremsstrahlung) in the Coulomb field of a nucleus~\cite{cd2,cd1}. In
the pion alab frame, the nuclear Coulomb field effectively provides a virtual photon
beam incident on a pion target at rest. At small invariant momentum transfer t$\leq
10^{-4}$ (GeV/c)$^2$, where t equals the photon mass square, the virtual photons are
quasi-real. In addition, the electromagnetic contribution to the scattering amplitude
is large compared to meson exchange contributions. This allows one to measure the
pion polarizability (a hadronic quantity) via a well understood QED probe.

The Primakoff differential cross section of this process in the alab frame may
be expressed as~\cite{starkov}:
\begin{equation}
\label{eq:Primakoff_1}
\frac
{{d}^3{\sigma}}
{{dt}{d}{\omega}{d\cos{\theta}}}
=
\frac
{\alpha_{f}{Z}^2}
{\pi\omega}
\cdot
\frac
{t-t_{0
}}
{t^2}
\cdot
\frac
{{d}\sigma_{\gamma\pi}{(}\omega,\theta{)}}
{{d}{\cos}{\theta}},
\end{equation}
\noindent
with the following expression for the $\gamma\pi$ cross section in the pion
alab frame:
\begin{equation}
\label{eq:Primakoff_2}
\frac
{{d}\sigma_{\gamma\pi}{(}\omega,\theta{)}}
{{d}{\cos}{\theta}}
=
\frac
{{2}{\pi}{\alpha_{f}}^2}
{{m}_{\pi}^2}
\cdot
\{
{F}_{\gamma\pi}^{pt}{(}{\theta}{)}
+
\frac
{{m_{\pi}}{\omega}^2}
{\alpha_{f}}
\cdot
\frac
{\bar{\alpha}_{\pi}{(}1+{\cos}^{2}{\theta}{)}+2\bar{\beta}_{\pi}{\cos\theta}}
{{(}{1+\frac{\omega}{m_{\pi}}{(1-\cos{\theta})}}{)}^3}
\}.
\end{equation}
\noindent
Here, t$_0=(m_\pi\omega/p_{b})^2$, with ${p_{b}}$ the incident pion beam momentum
in the laboratory, $\theta$ is the scattering angle of the real photon relative to
the incident virtual photon direction in the alab frame, $\omega$ is the energy of
the virtual photon in the alab frame, $Z$ is the nuclear charge, $m_\pi$ is the pion
mass, $\alpha_{f}$ is the fine structure constant and $\bar{\alpha_\pi}$,
$\bar{\beta_\pi}$ are the pion polarizabilities. The energy of the incident virtual
photon in the alab (pion rest) frame is:
\begin{equation}
\omega \sim  (s - {m_{\pi}}^2)/2m_{\pi};
\label{eq:omega}
\end{equation}
\noindent
so that the energy of the photon is determined by s, the squared mass of the
$\gamma\pi$ final state. The function ${F}_{\gamma\pi}^{pt}{(}{\theta}{)}$ accounts
for the angular dependence of the point Thomson cross section, for scattering in a
pure Coulomb field. For pion scattering, it reads:
\begin{equation}
\label{eq:Primakoff_3}
{F}_{\gamma\pi}^{pt}{(}{\theta}{)}=
\frac{1}{2}\cdot
\frac
{1+{\cos}^{2}{\theta}}
{{(}{1+\frac{\omega}{m_{\pi}}{(1-\cos{\theta})}}{)}^2}
{.}
\end{equation}

From Eq.~\ref{eq:Primakoff_2}, the cross section depends on
$(\bar{\alpha}_{\pi}+\bar{\beta}_{\pi})$ at small $\theta$, and on
$(\bar{\alpha}_{\pi}-\bar{\beta}_{\pi})$ at large $\theta$. A precise fit of the
theoretical cross section (Eq.~\ref{eq:Primakoff_1}-\ref{eq:Primakoff_3}) to the
measured angular distribution of scattered photons, allows one to extract the pion
electric and magnetic polarizabilities. Fits will be done for different regions of
$\omega$ for better understanding of the systematic uncertainties. We will carry out
analyses with and without the dispersion sum rule constraint that
$\bar{\alpha}_{\pi}+\bar{\beta}_{\pi}\approx0.4$. We can achieve a significantly
smaller uncertainty for the polarizability by including this constraint in the fits.
For fits without this constraint, the statistics requirement is a factor of 10-100
higher. Such unconstrained fits will also be of even greater interest.

The event generator produces events in the alab frame, characterized by three
kinematical variables, t, $\omega$ and $\cos(\theta)$, and distributed with a
probability, given by the theoretical Compton-Primakoff cross section
(Eq.~\ref{eq:Primakoff_1}-\ref{eq:Primakoff_3}). Then, the photon-pion scattering
kinematics are calculated. The virtual photon, specified by its four-vector
components $k=\{\omega,\vec{k}\}$ and squared mass t=M$^2=\omega^2-|\vec{k}|^2$,
incident along the recoil direction $\vec{k}/|k|$, is scattered on the pion "target"
and emerges as a real photon with energy $\omega^{\prime}$ at an angle $\theta$:
\begin{equation}
\label{eq:Compton}
{\large
\omega^{\prime}=\frac
{\omega{(}{1}+\frac
{\omega^2{-}|\vec{k}|^2}{{2}{m}_{\pi}{\omega}}
{)}}
{
{1}{+}
\frac{\omega}{{m}_{\pi}}
{(}{1-}
\frac{|\vec{k}|}{\omega}
\cos\theta
{)}
}
}
\end{equation}

The photon azimuthal angle around the recoil direction is randomly generated using a
uniform distribution. Afterwards, all four-vector components of all reaction
participants (pion, photon and recoil nucleus) are calculated in the alab frame. The
azimuthal angle of the recoil nucleus is also randomly generated by a uniform
distribution. Finally, the reaction kinematics are transformed to the CM and then to
the lab frame by a Lorentz boost.

\begin{figure}[tbc]
\centerline{\epsfig{file=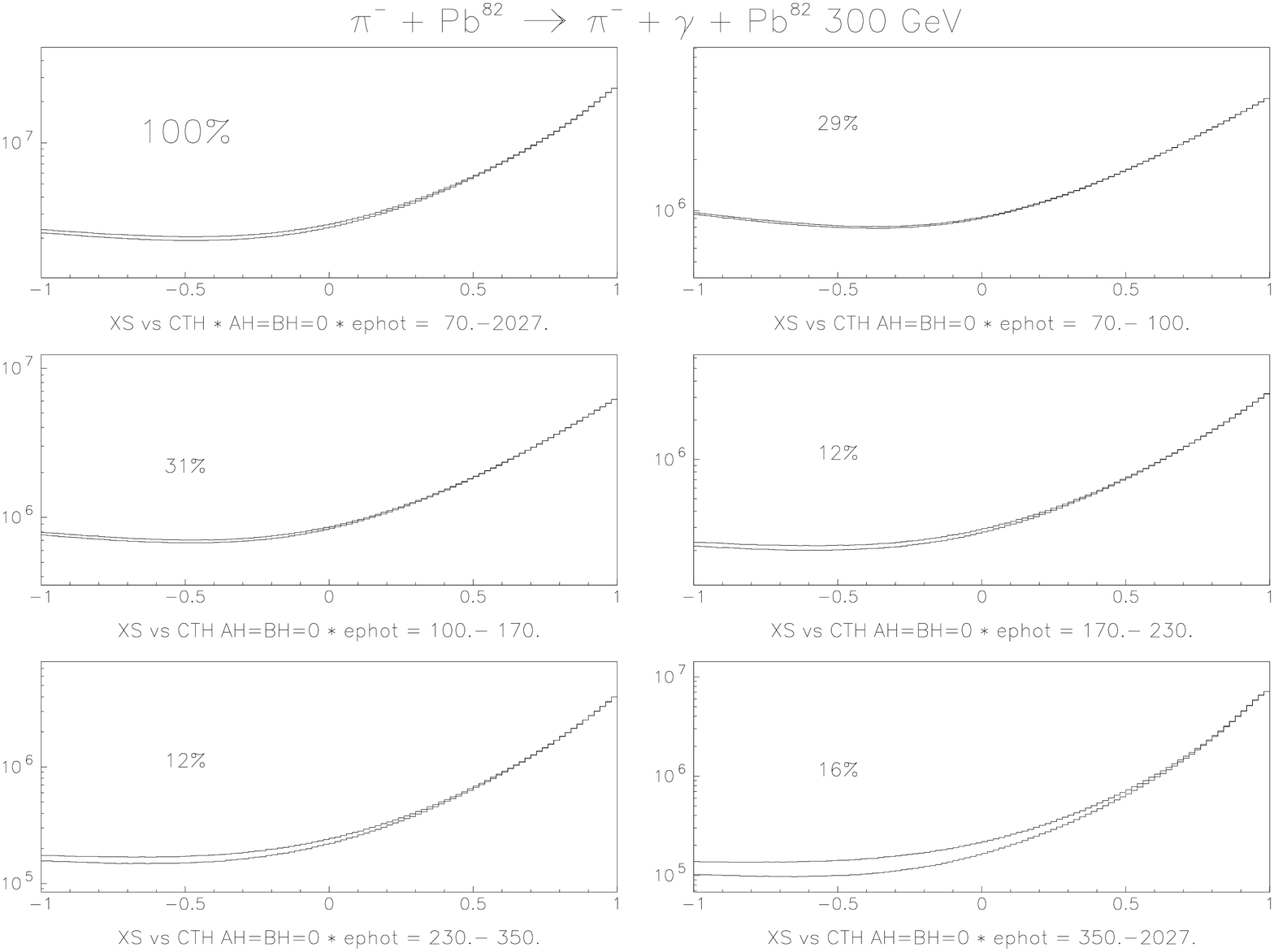,width=16cm,height=10cm}}
\caption{The dependence of the theoretical angular distributions on
polarizability for different regions of $\gamma$ energy $\omega$ 
(given in MeV), function of $\cos(\theta)$ in the alab frame.
The lower curve corresponds to $\bar{\alpha}$=7,
$\bar{\beta}=-$6; while the upper curve corresponds
to zero polarizabilities. The percentage shows the statistics fraction
in each $\omega$ region.}
\label{fig:cth_sens_1}
\end{figure}

\begin{figure}[tbc]
\centerline{\epsfig{file=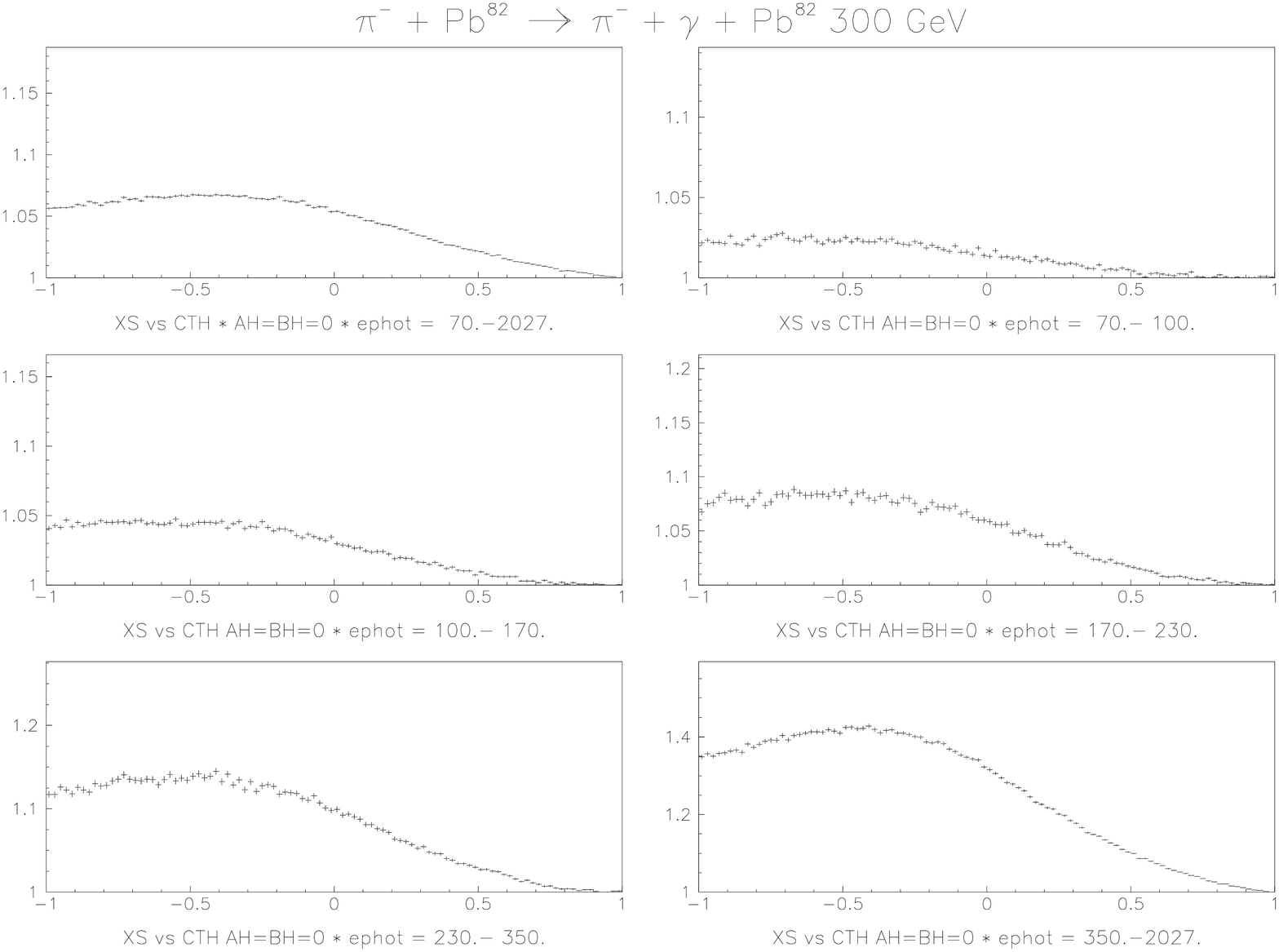,width=16cm,height=10cm}}
\caption{Ratio of the theoretical angular distributions for different
regions of $\gamma$ energy $\omega$ (given in MeV), as a function of
$\cos(\theta)$ in the alab frame, for
the case of zero polariabilities (Thomson term only),
relative to the case in which
$\bar{\alpha}$ =7, $\bar{\beta}=-$6.
The contribution of the polarizability to the
cross section is larger at back alab angles, and increases with
increasing $\omega$.}
\label{fig:cth_sens_3}
\end{figure}

For the measurement of the pion electric ($\bar{\alpha}_{\pi}$) and magnetic
($\bar{\beta}_{\pi}$) polarizabilies, one must fit the theoretical cross section
(Eq. \ref{eq:Primakoff_1}-\ref{eq:Primakoff_3}) to measured distributions, after
correcting for acceptance losses. The dependence of the theoretical angular
distributions on polarizability  (for $\bar{\alpha}$ =0, ~6.8) for different regions
of $\gamma$ energy $\omega$ in the alab frame is given in
Fig.~\ref{fig:cth_sens_1}-\ref{fig:cth_sens_3}. The sensitivity to the polarizability
increases with increasing $\omega$ energy and at back angles. We carried out several
tests (a) by comparing the generated one-dimensional t, $\omega$ and $\cos(\theta)$
distributions (integrated over the other two variables) with the corresponding
theoretical cross sections, (b) by comparing the Monte-Carlo computed total cross
section with the theoretical value, given by numerical integration of
Eq.~\ref{eq:Primakoff_1}-\ref{eq:Primakoff_3}, (c) by fitting the generated
$\cos(\theta)$ event distribution with the theoretical cross section, and getting back 
the input values of $\bar{\alpha_\pi}$ and $\bar{\beta_\pi}$. A convenient method is
to use the $\cos(\theta)$ distribution integrated over t and $\omega$, which is most
sensitive to the polarizability effect. We performed fits to this distribution in
different $\omega$ regions, and for the entire $\omega$ domain. For the fit, we used
a MINUIT routine, which minimizes the $\chi^2$ statistic between the theoretical and
measured points, with three free parameters: $\bar{\alpha}_{\pi}$, $\bar{\beta}_{\pi}$
and a normalization constant. The routine evaluates those parameters, as well as
their statistical error. These tests ensured us that the simulated distributions are
correct to a high degree of precision, as needed for the measurement of the
polarizability with $\Delta\bar{\alpha}_{\pi}\approx0.2$ .

\subsubsection{Design of the Primakoff Trigger\label{sec:how}}

The small Primakoff cross section and the high statistics required for extracting the
hadron polarizability requires a data run at high beam intensities with good
acceptance. This sets the main requirements for the trigger system:
\begin{itemize}
\item it has to act as a "beam killer" , to suppress the high rate background
associated with non-interacting beam pions
\item it has to avoid cutting the
  acceptance at the important photon back angles in the alab frame,
where the hadron polarizability measurement is most sensitive.
\item it has to cope with background from low energy $\gamma$'s
or delta electrons caused by  the beam passing through the apparatus.
\end{itemize}

We want to adapt a Primakoff trigger by a veto of the unscattered beam in a window on
the hadron energy and a coincidence of the scattered pion with a $\gamma$ measured in
the calorimeter.

To study the feasibility of such a trigger scenario, simulations were carried out at
300 GeV, for illustrative purposes. The actual measurements will most likely be
carried out at 280 GeV.

\begin{figure}[tbc]
\centerline{\epsfig{file=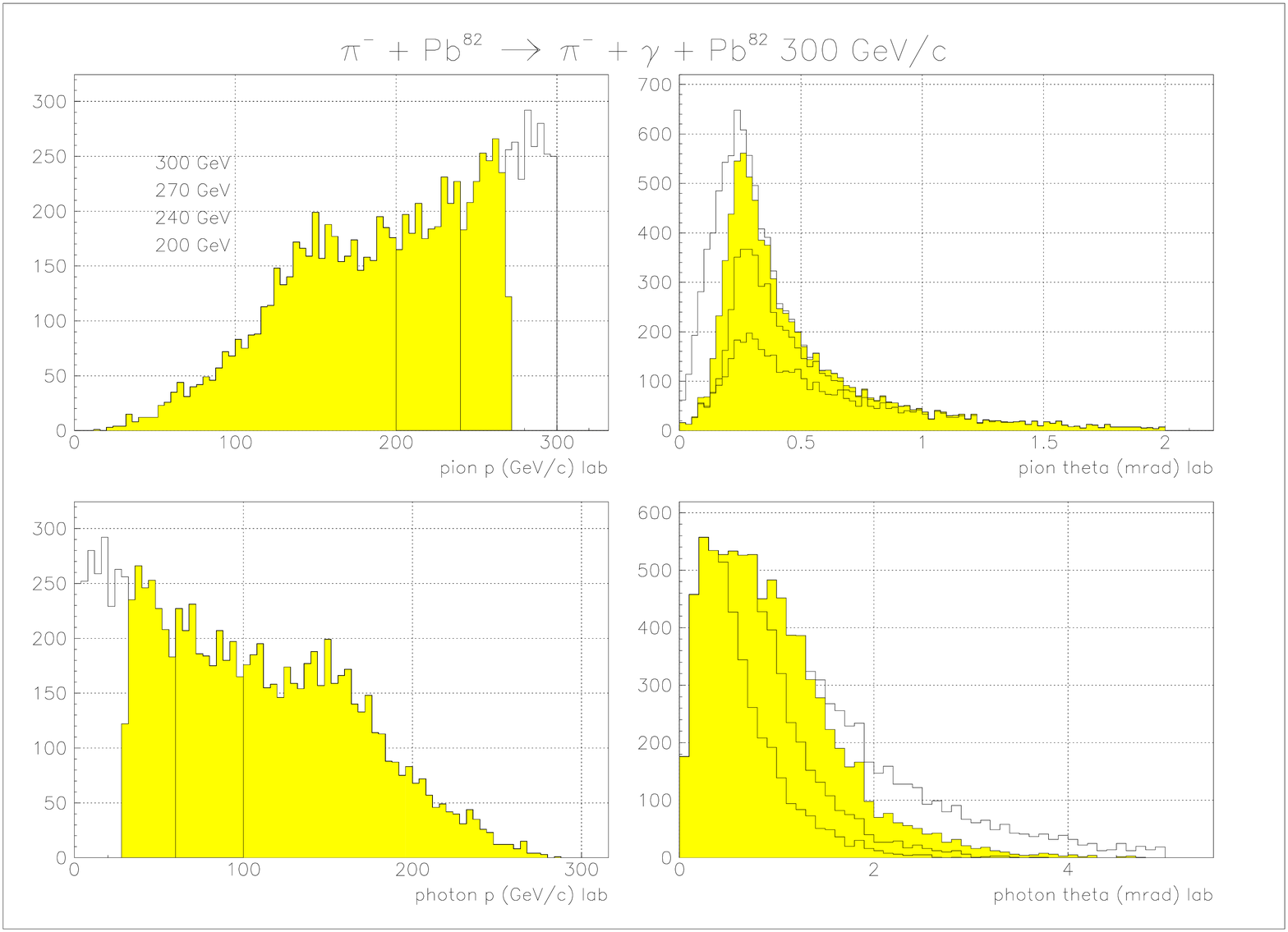,width=16cm,height=10cm}}
\caption{MC simulation showing the kinematics of the
$\pi\gamma\rightarrow\pi\gamma$ reaction, in terms of the $\pi$ and $\gamma$
momenta and angles. The overlayed spectra correspond to different trigger cuts
on the final state $\pi$ momentum.}
\label{fig:kinematics}
\end{figure}

\begin{figure}[tbc]
\centerline{\epsfig{file=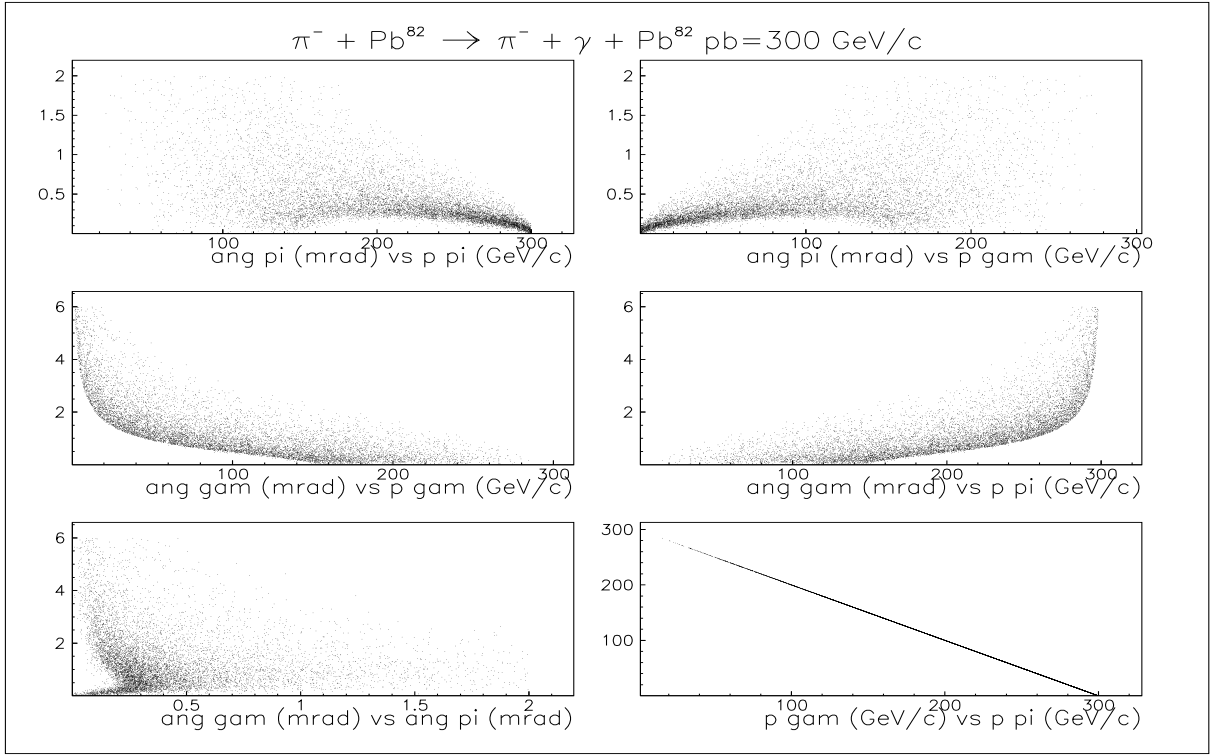,
width=16cm,height=10cm}}
\caption{MC simulation showing the correlation between the $\pi$ and
$\gamma$ kinematic variables in the lab frame.}
\label{fig:pi_gamma_correl}
\end{figure}

\begin{figure}[tbc]
\centerline{\epsfig{file=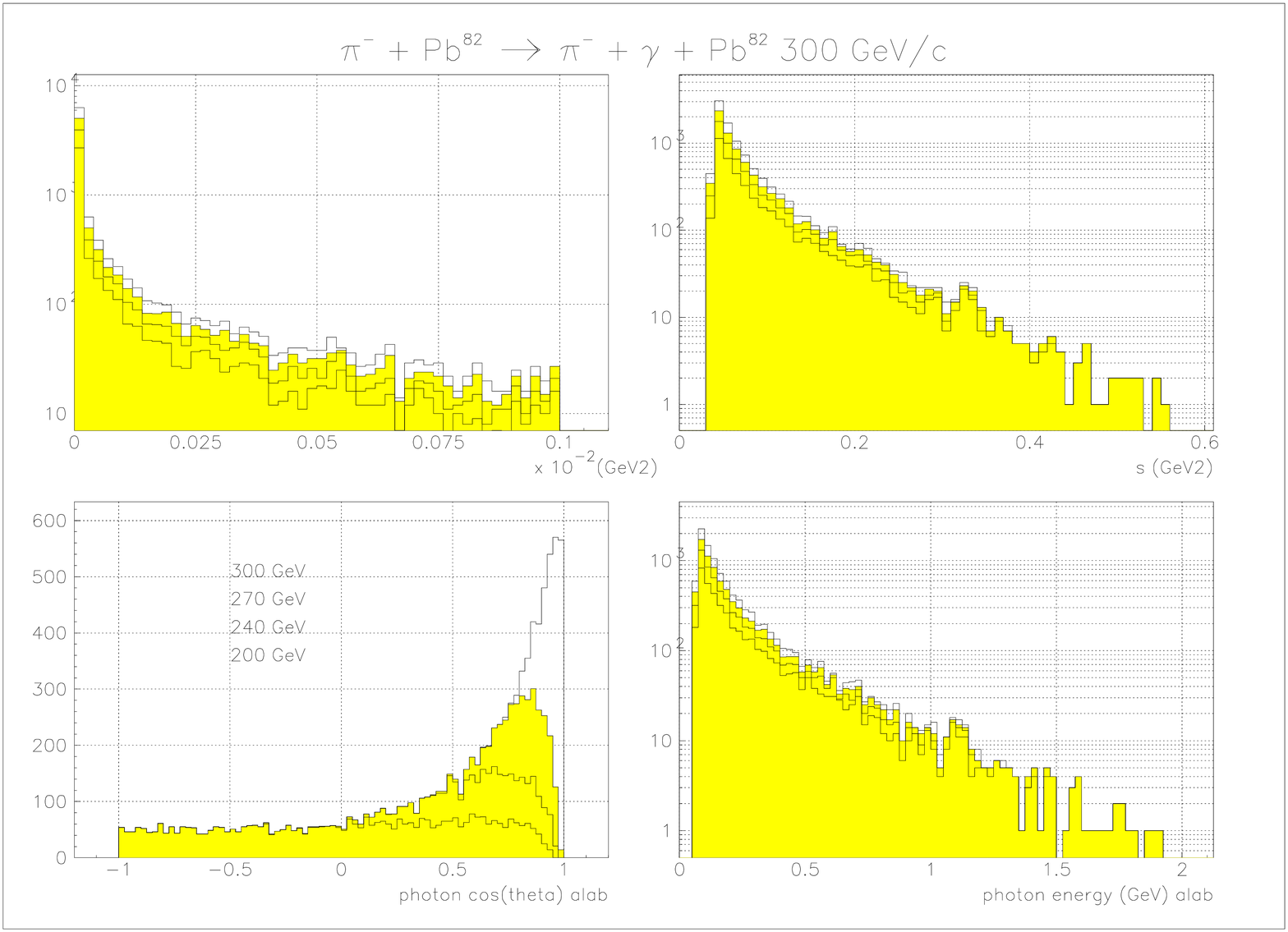,width=16cm,
height=10cm}}
\caption{MC simulation showing the acceptance of the
$\pi\gamma\rightarrow\pi\gamma$ reaction in terms of the invariant four
momentum transfer t to the target,the squared invariant energy s of the final
state $\pi\gamma$, the angular distribution versus $\cos(\theta)$ with
$\theta$ the gamma scattering angle in the alab frame, and the virtual photon
energy in the alab frame. The overlayed spectra correspond to different cuts
on the final state $\pi$ momentum.}
\label{fig:acceptance}
\end{figure}

For the reaction given in Eq. ~\ref{eq:polariz} (at 300 GeV/c), the laboratory
outgoing $\gamma$'s are emitted within an angular cone of up to 5 mrad, and the
corresponding outgoing $\pi$'s are emitted up to 2 mrad. The $\gamma$ energies range
from $0-280$ GeV, and the corresponding outgoing $\pi$ energies range from $20-300$
GeV. This is shown in the
Figures~\ref{fig:kinematics},~\ref{fig:acceptance},~\ref{fig:pi_gamma_correl}
and~\ref{fig:cth_correl}. We consider the Compton scattering angular distribution in
the alab frame. The recoil nucleus of mass M$_T$ for a Primakoff reaction has
negligible recoil energy (T$_r~\approx~$t/2 M$_T$), with roughly 99\% of the generated
events having  target recoil kinetic energies less than 30 keV. Therefore, the final
state $\pi$ and $\gamma$ effectively carry all the four momentum of the beam pion, so
that momentum and energy conservation may be used at the trigger level for background
suppression.

For a 300 GeV pion beam, our Monte Carlo simulations (see Fig.~\ref{fig:cth_correl})
show that we lose very little polarizability information by applying an "energy cut"
trigger condition that rejects events in which the final state charged pion has more
than 240 GeV, and the final state $\gamma$ has less than 60 GeV. The 240 GeV cut
value was devised to act as a beam killer, as discussed in more detail below. The 60
GeV cut will also be very effective in reducing the $\gamma$ detector (ECAL2) trigger
rate, since a large part of the background $\gamma$ rate for a 300 GeV beam energy is
below 60 GeV.

The polarizability insensitivity to these cuts results from the fact that the most
forward (in alab frame) Compton scattering angles have the lowest laboratory $\gamma$
energies and largest laboratory angles. In addition, the cross section in this
forward alab angle range is much less sensitive to the polarizabilities. This is seen
from Eq. \ref{eq:Primakoff_3}, since with
$\bar{\alpha}_{\pi}+\bar{\beta}_{\pi}\approx 0$ from the dispersion sum rule, the
polarizability component is small at forward compared with the back angles. The
acceptance is reduced by the energy cut for the forward alab angles (shown in
Fig.~\ref{fig:acceptance} for the alab frame), but is unaffected at the important
alab back angles. In practice, the acceptance for alab cos($\theta) < -0.9$ will be
reduced in off-line analysis, since this angular range corresponds to laboratory
outgoing pion angles less than 100 $\mu$rad. Such events will be rejected since their
 z position and momenta cannot be well determined in part due to the 40 $\mu$rad
angle measurement error from the Coulomb multiple scattering. However, the number of
such events is limited, and their exclusion from the final fits should not
significantly affect the polarizability determination. Summarizing, the purposes of
the pion and $\gamma$ energy constraints at the trigger level are fulfills the "beam
killer" requirement and at the same time removes backgrounds associated with low
energy $\gamma$'s  or delta electrons or e$^+$e$^-$ pairs incident on ECAL2,

Alternatively one might consider a trigger scenario based on a measurement of the
track angles closed to the target. The SELEX/E781 \cite {selex} experiment at
Fermilab used a mixed pion/hyperon beam at 600 GeV. The Primakoff physics in this
experiment was attempted as a parasitic experiment, relative to the main charm topic.
The charm trigger constraints did not allow implementing a $\gamma\pi$ coincidence
condition for the first level trigger. SELEX therefore implemented  a fast first
level T1 "beam kill" Primakoff trigger, which did not include a signal from the gamma
detector. Instead, the trigger worked by distinguishing pions with a small Primakoff
scattering angle from  non-interacting straight through beam pions. This trigger
employed  six (H,V) silicon planes grouped in 3 stations, two before and one after the
target, with event-by-event readout. A processor provided a fast trigger for a
scattering angle greater than 150 $\mu$rad  at 250 nsec after beam crossing. In order
to further reduce the DAQ rate, another trigger condition was required at the later
T2 level. For this purpose, SELEX used a minimal energy deposit trigger of 100 GeV
from the photon detector located 50 meters downstream. It was built by electronically
summing up and then applying a discriminator to the signals of about 300 PbG blocks.

\begin{figure}[tbc]
\centerline{\epsfig{file=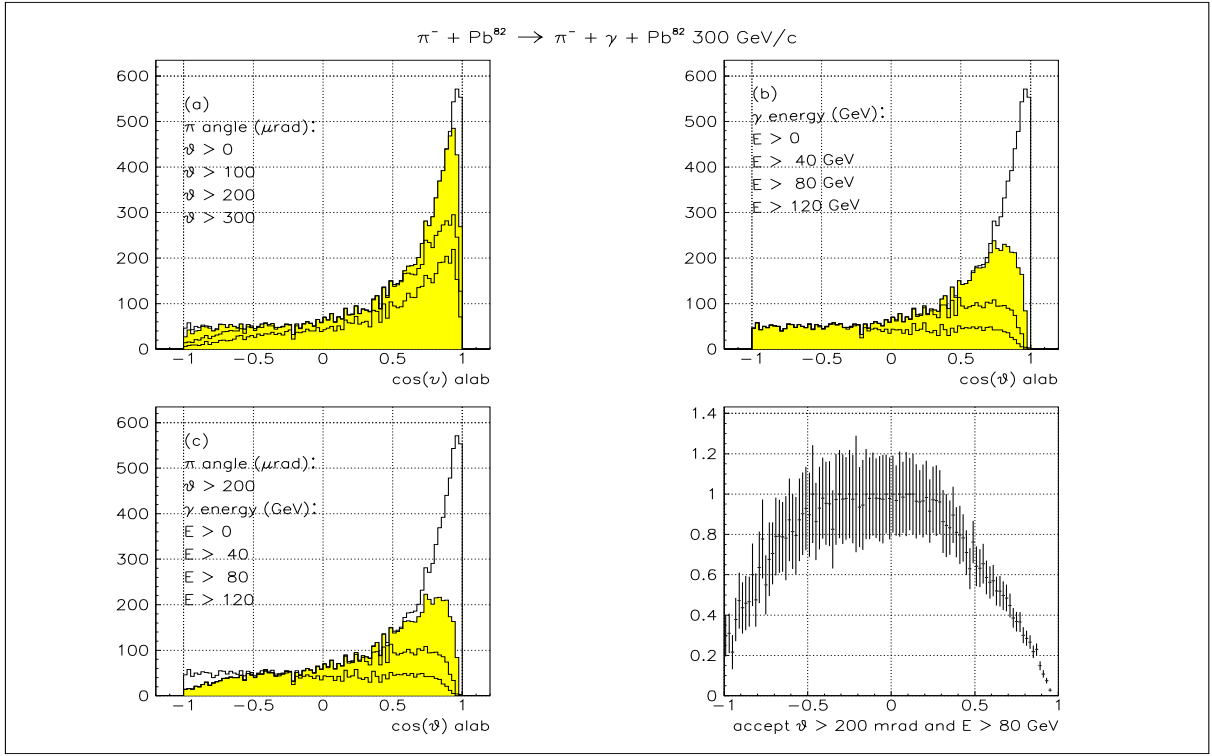,width=16cm,height=10cm}}
\caption{Monte Carlo simulation showing the acceptance in the photon
scattering angle $\cos(\theta)$ in alab frame for: (a) cuts on the pion
lab scattering angle, (b) cuts on the gamma lab energy, and (c) a
combination of (a) and (b). The last plot shown the acceptance after these cuts.
The $\cos(\theta)$ range in which the acceptance is essentialy equal to 1 is very 
limited.}
\label{fig:cth_acc}
\end{figure}

As seen from the simulation in Fig.~\ref{fig:cth_acc}, the scatter angle trigger
technique cuts the acceptance in the back angle range. Therefore, acceptance
corrections and their uncertainties would be required for analysis of data taken with
such a trigger. SELEX to this date was only able to have limited statistics dedicated
data runs for the $\pi\gamma$, $\pi\pi^0$, $\pi\eta$ Primakoff final states.

\begin{figure}[tbc]
\centerline{\epsfig{file=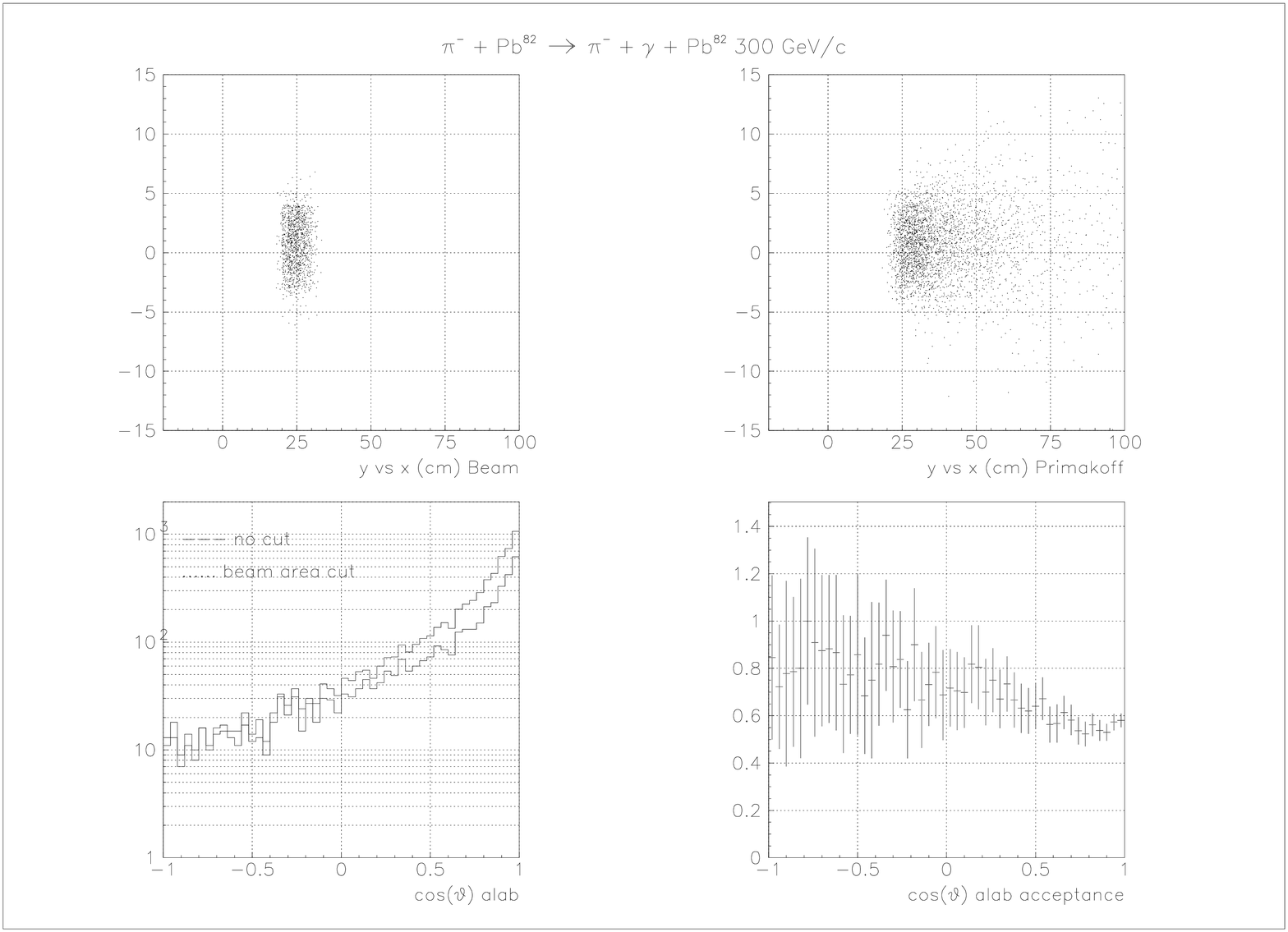,width=16cm,height=10cm}}
\caption{Monte Carlo simulation showing the acceptance in the photon
scattering angle $\cos(\theta)$ in alab frame in the case of a cut
on the beam (x,y) spot at the photon detector position. Upper left:
pure beam spot with magnetic deflection (without target). Upper right:
area covered by Primakoff events. Lower left: $\cos(\theta)$ distribution
before and after a (x,y) cut on the beam spot. Lower right: $\cos(\theta)$
acceptance as ratio of the two spectra in the lower left plot. See the text
for more details.}
\label{fig:cth_beam_cut}
\end{figure}

 We also considered a Primakoff trigger solution for COMPASS based on an position cut
on the unscattered beam spot, at the maximal possible distance downstream, at the
photon detector position. We carried out a Monte Carlo simulation in which we
generated Primakoff events using the beam phase space from the Fermilab SELEX beam.
The COMPASS beam properties given in Table~\ref{tab:beam} are somewhat better. The
SELEX  beam had a spot size on target of about 1 cm$^2$ and a divergence of about 1
mrad. The divergence is thereby larger than the average Primakoff angle, which
literary places the Primakoff events "in the beam". We took into account the COMPASS
magnetic deflection and considered a cut of about 8$\times$8 cm$^2$  over the beam
spot at 40 meters downstream from target, where the COMPASS photon detector may be
positioned. In the Fig.~\ref{fig:cth_beam_cut} upper plots, we compare the beam spot
of the unscattered beam with the area covered by Primakoff events generated with the
same beam phase space. We note the large overlay, which results in a considerable cut
on Primakoff events. In the lower plots we show the photon alab $\cos(\theta)$
distribution before and after this cut. We note the important acceptance cut, due to
alab back scattered photons hitting the beam veto area, which (as in the case above)
would result in uncertain acceptance corrections.

\subsection{Beam Requirements}

The beam requirements for COMPASS Primakoff runs are given in Table~\ref{tab:beam}.
Two beam Cherenkov detectors (CEDARS) far upstream of the target provide $\pi/K/p$
PID. The incoming hadron momentum is measured in the beam spectrometer. Before and
after the target, charged particles are tracked by high resolution silicon strip
tracking detectors. The final state hadron and $\gamma$ momenta are measured
downstream in the magnet and in the photon calorimeter, respectively.  This allows a
precise determination of the small p$_T$ kick to the target, the main signature of
the Primakoff process, and the means to separate Primakoff from diffractive
scattering.

The measurement of both initial and final state momenta provides constraints to
identify the reaction. Since beam particles are identified in the CEDARS, and since
we study simple exclusive reactions, there is no need for PID in the final state via
a RICH. We can get quality statistics for the pion  study with beam intensities of 5
MHz (1.25 $\times$ 10$^7$ particles in a 2.5 second spill ) Some of the detectors
(such as HCAL2 modules with a signal duration of about 50 nsec) needed for this study
must accept the full beam intensity, and cannot tolerate beam intensities higher than
5 MHz. Beam rates lower than the final COMPASS rates are planned for this study, in
which many of the COMPASS systems (DAQ, detectors, etc.) must be implemented. In this
way, we proceed in a staged approach to study also the problems (dead time, pile up,
radiation damage, etc.) associated with running with design beam intensities as high
as 40 MHz. With the lower beam rates planned, we should also be able to achieve
better beam quality. For example, we  describe later that the electromagnetic
calorimeter ECAL2 will have a hole in it (vertical size 7.6 cm, or 2 GAMS blocks) to
allow all of the non-interacting beam and the main part of the Primakoff scattered
pions (emitted at small laboratory angles) to pass through towards the hadron
calorimeter HCAL2 located behind ECAL2. The beam should be tuned/collimated to be
centered in this hole, with minimum halos. It is also important that data are taken
with different beam energies and targets, as part of an effort to control systematic
errors. Data should also be taken with both positive and negative beams (including
proton beam).

\begin{table}
\begin{center}
\begin{tabular}[tbc]{|l|l|}
\hline
Beam momentum (average)          & 300 GeV/c (or lower)\\
Momentum spread (rms)            &  13 GeV/c \\
Momentum resolution (rms)        &  2 GeV/c  \\
Beam cycle                       & extraction 2.5 sec, total cycle 14.8 sec   \\
Beam angular divergence (rms)    & $\sigma_V=0.5$ mrad, $\sigma_H=0.5$ mrad \\
Beam particles                   & $\pi^{\pm}$, K, p                \\
Beam particle ID                 & 2 CEDARS (Cherenkov) in beam line    \\
Beam intensity  (part./spill) & 1.25~10$^7$ \\
\hline
\end{tabular}
\end{center}
\caption{Beam settings for Primakoff measurements.}
\label{tab:beam}
\end{table}

\subsection{Target and Target Detectors}

The main Primakoff target will be Pb which has a 1.2 mb Compton scattering
(polarizability) cross section and total inelastic cross section of 1.8 barn. We also
need Primakoff scattering on nuclei with Z$<82$ to see the Z$^2$ cross section
dependence, and to make sure experimentally that there are no higher order
corrections for atomic numbers as high as Z=82. We plan to arrange these targets
along the beamline with a spacial distance large enough, so that the target with the
interaction can clearly be identified from the position of the kink in the  pion
trajectory (change in x,y slopes between beam and final state pion tracks)

With a 1\% interaction length target, we expect roughly 80 $\pi\gamma$ events/spill
for a beam of 2.5~10$^7$ pions/spill. Each Primakoff target is followed by two
interaction counters (IC) with a triggering condition of 1 MIP each. We will check
offline (not at the trigger level) that the detectors downstream of the targets (at
different z-positions) have one charged particle. We need Si tracking detectors
before and immediately after the targets. We veto target break-up events by selecting
 1 MIP in the scintillation IC interaction counters after the targets, and by
selecting low-t events in the off-line analysis. The target z-resolution (position)
is less/more than $\pm$ 15 cm rms for events with outgoing pion scattering angles
more/less than 100 $\mu$rad.

\subsection{The Magnetic Spectrometer and the t-Resolution.}

We need good resolutions in momentum for the incident and final state pions and
$\gamma$'s. Tracking before and after the magnet is required. In this way, the
important four momentum resolution {\it t} (momentum transfer to the target nucleus)
can be kept as good as possible. A final state $\pi^-$ at 200 GeV/c can be momentum
analyzed to 2 GeV/c resolution, with better resolution at lower momenta.

The angular resolution for the final state $\pi$ should be good, which we may achieve
by minimizing the multiple scattering in the targets and detectors. With a lead
target of 1\% interaction length = 2 g/cm$^2$ =0.30 radiation length, the beam and
outgoing pion multiple Coulomb scattering in the target gives an rms angular
resolution of 40 $\mu$rad. For this estimate, we assumed that a Primakoff 150 GeV
pion is produced at the center of the target, and we added in quadrature the 18
$\mu$rad contribution of the incident 300 GeV pion and the 35 $\mu$rad.  contribution
of the outgoing pion. We expect to have angular resolution not counting multiple
scattering of about 6 $\mu$rads using for example 20 $\mu$m Si strips. This is based
on an expected position resolution of 3 $\mu$m  (using cluster centroid), and vertex
planes 50 cm downstream from target. Thus, the vertex detector angular resolution is
significantly better than the multiple scattering contribution to the angular
resolution. We estimate the resolution of the transverse momentum p$_T$ considering
the p$_T$ generated through MCS for a straight-through beam pion of 200 GeV. The
p$_T$ given to such a beam pion no Compton scattering) is then $p_T = p \times
\Delta{\theta} = 200 \times \times 10^{-6} = 8$ MeV, which corresponds
t~=~p$_T^2$~=~0.6~$\times~10^{-4}$ GeV$^2$. Other contributions to the t-resolution are
the uncertainties in beam angle and momentum, detector material and air downstream of
the target, and energy/position resolution in ECAL2. All together, we aim for a p$_T$
resolution less than 15 MeV, corresponding to $\Delta t$ better than $\approx$ 2.5
$\times 10^{-4}$ GeV$^2$ over the energy range 40-240 GeV. These considerations fix the
inherrent $\Delta$t in determining t = p$_T^2$ for Primakoff Compton scattering.

Our resolution goal is based on the need for an effective t-cut to minimize
contributions to the Coulomb data from diffractive production. We require a
t-resolution that is about a factor of 10 smaller than the slope in t observed for
diffractive data on a Pb target. The Pb diffractive data falls as $\exp(-t/0.0025)$ with
t expressed in GeV$^2$. This means that we need resolution $\Delta$t=0.00025 GeV$^2$,
or a resolution in transverse momentum less than 15 MeV. This conservative goal is
based on the t distributions measured at a 200 GeV low statistics but high resolution
experiment for $\pi^- \rightarrow \pi^- \pi^0$ \cite {jens} and $\pi^- \rightarrow
\pi^- \gamma$ \cite {pigam} Primakoff scattering at 200 GeV at FNAL. The t
distribution of the $\pi^- \rightarrow \pi^- \gamma$ \cite {pigam} data agrees well
with the Primakoff formalism out to t~=~$10^{-3}$ GeV$^2$, which indicates that the
data is indeed dominated by Coulomb production.

In practice, we will study the t resolution and backgrounds in initial data runs. As
part of our systematics studies, we plan to take some data with 0.5\% rather than 1\%
interaction length Pb target. Minimum material (radiation and interaction lengths) in
COMPASS will also give a higher acceptance, since that allows $\gamma$'s to arrive at
ECAL2 with minimum interaction losses, while producing minimum $e^+e^-$ backgrounds.
That is, the fully instrumented COMPASS spectrometer is not needed. Only minimum
equipment (the bare bones) should be used for this Primakoff physics, which also
matches the constraints of a limited COMPASS budget.

\subsection{The Photon Calorimeter ECAL2}

\begin{figure}[tbc]
\centerline{\epsfig{file=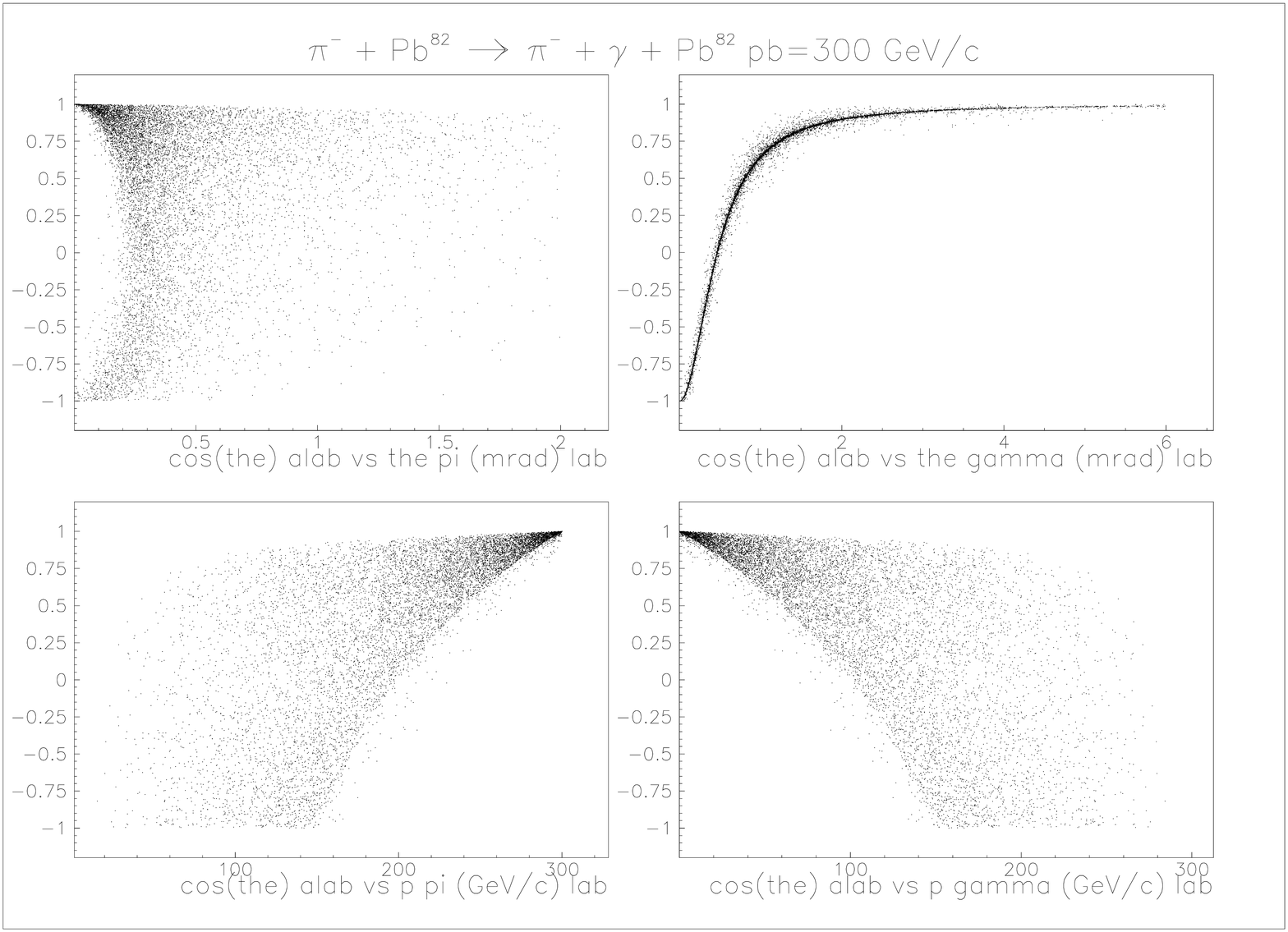,width=16cm,height=10cm}}
\caption{MC simulation showing the correlation between the $\cos(\theta)$
of the virtual photon in alab frame and the angles (upper plots) and
momenta (lower plots) of the $\pi$ and $\gamma$ in the lab frame.}
\label{fig:cth_correl}
\end{figure}

The discussions for ECAL2 and HCAL2 follows our simulations, which  were carried out
at 300 GeV for illustrative purposes. The actual measurements will most likely be
carried out at 280 GeV. In COMPASS, we can measure a final state 200 GeV $\gamma$ to
$\pm$2 GeV, with a position resolution of 1.5 mm, in the second photon calorimeter
ECAL2. We plan to use an ECAL2 $\gamma$ detector equipped with 3.8 by 3.8 cm$^2$
GAMS-4000 lead glass blocks to make a total circular active area of order 1.5 m
diameter. The GAMS-4000 blocks are adequate. This is so, since radiation damage will
be negligible for this run with beam intensity of at most 2.5 $\times$ 10$^7$ per
spill. The hybrid meson study ($\eta$ detection) sets the area of ECAL2.

The p$_T$ kicks of the two COMPASS magnets are 0.45 GeV/c for SM1 (located 4 meters
from target) and 1.2 GeV/c for SM2 (located 16 meters from target). We require the
highest conveniently accessible effective p$_T$ kick for this physics. The fields of
both magnets must therefore be set $\it{additive}$ for maximum deflection of the beam
from the zero degree (neutral ray) line. We need to maximize the distance from the
zero degree line to the beam hole in ECAL2 (located 40 meters or more from target),
to attain at least 10 cm for the distance between the zero degree line and the hole
edge. This is so since the Primakoff $\gamma$'s are concentrated around the zero
degree line (see Fig.~\ref{fig:cth_correl}), and a good $\gamma$ measurement requires
clean signals from 9 blocks, centered on the hit block. The blocks near zero degrees
should be selected to be the ones with the very best performance, and they should
have accurate gain monitoring. ECAL2 should be at maximum distance from the target
(we assume 40 meters) to also maximize the distance between the zero degree line and
the deflected beam position at ECAL2. The hole size and position must be optimized to
minimize the hadrons hitting ECAL2 blocks at the hole perimeter. We plan it to be big
enough (2 blocks V $\times$ 16 blocks H) to pass completely the non-interacting beam,
and to pass also the majority of Primakoff scattered pions. In that way, these
particles are measured well in the HCAL2 hadron calorimeter behind ECAL2. We are then
able to optimally fix the beam killer threshold cut. We may also better understand
the detector behavior by comparing the energy determinations of tracking detectors
and HCAL2, for those pions arriving at HCAL2 without traversing ECAL2.

Besides the polarizability $\pi\gamma$ detection, COMPASS should also detect
$\pi\pi^0$ for the anomaly study and $\pi\eta$ for the hybrid study. The two
$\gamma$'s from $\pi^0$ and $\eta$ decay have opening angles $\theta_{\gamma\gamma}$
for the symmetric decays of $\theta_{\gamma\gamma}= m/4E_t$, where m is the mass
($\pi^0$ or $\eta$) and E$_t$ is the lowest energy $\pi^0$ or $\eta$ to be detected.
And the opening angles are increased for the asymmetric decays. The $\pi^0$'s and
$\eta$'s themselves are produced with an angular spread around the beam direction.
The consequence is that to obtain good acceptance for all the Primakoff reactions
discussed above, an ECAL2 with size of order 1.5 m diameter is required. This may be
achieved with GAMS-4000 lead glass blocks, each of size $3.8\times3.8$ cm$^2$.

The $300\pm13$ GeV unscattered beam should be measured with HCAL2 to be between
240-360 GeV with 99\% probability. From MC simulations, the number of Primakoff
scattered pions below 40 GeV is less than 0.3\%, so that 40 GeV pions are about the
lowest energy of interest. In any case, too low energy pions may be blocked by the
magnet yoke. We will effectively set a $\pi^-$ acceptance energy window of 40 - 240
GeV, via a minimum threshold of  60 GeV for the $\gamma$ energy deposited in ECAL2,
and an HCAL2 veto for energies above 240 GeV. The size of the BP trigger
scintillation detector (see section ~\ref{sec:trigger} below) must be of order
$60\times15$ cm$^2$ to accept unscattered beam and Primakoff scattered events. The
vertical size of this BP detector is larger than the 10 cm needed for polarizability.
It must match the anomaly and hybrid trigger, to catch also the scattered $\pi^-$'s
associated with the $\pi^-\pi^0$ and $\pi^-\eta$ final states. Monte Carlo studies
are in progress to optimize the vertical size of BP. The ECAL2 blocks will have their
gains well matched, and their analog signals will be electronically summed and
discriminated to provide a trigger signal on minimal energy deposit. In
table~\ref{tab:rate} we show the ECAL2 contribution to the Primakoff trigger.

Primakoff physics requires a very good energy resolution of photon calorimeters. For
the precise monitoring of the energy calibration of the photon calorimeters, COMPASS
may use a dedicated laser system, which was built by the Tel-Aviv University group
~\cite{steiner}.

\subsection{The Hadron Calorimeter HCAL2}

We intend to use beam rates of order 5 MHz where the rate limit is the maximum
allowed for good operation of the existing and tested 20$\times$20 cm$^2$ or 15
$\times$ 15 cm$^2$ Dubna hadron calorimeter  modules. For the beam killer trigger
purposes, we require a mini-HCAL2 configured  as an array of 15 $\times$ 15 cm$^2$
blocks (2 $\times$ 2 or  3 $\times$ 3) to catch non-interacting beam pions. The
energy sum for trigger purposes would be taken from this mini-HCAL2. However, we may
use a larger HCAL2 array (matching the ECAL2 size) as an aid in PID and in fixing the
HCAL2 threshold and as a check of the momentum determination by the tracking
detectors. Such a large HCAL2 can give further understanding of events where a hadron
hits ECAL2, and also for those that do not. The HCAL2 modules have energy resolution
of $\pm$15 GeV at 300 GeV. Together with the beam acceptance of $\pm$ 13 GeV, we can
achieve a 1-$\sigma$ identification of the beam via a detection window of 300 $\pm$ 20
GeV. We can therefore set a 3-$\sigma$ discriminator veto threshold at
$300-3\times20=240$ GeV, to veto 99\% of the beam. We will reduce the beam acceptance
to 13 GeV rms or lower, by collimation. With a lower threshold acceptance on  HCAL2
(say 220 GeV), we may achieve a yet higher beam rejection. The final value of the
energy cut will be set following in-beam tests. Here we just estimate that we will use
HCAL2 to reject events with with pion energies above 240 GeV for beam suppression. In
table~\ref{tab:rate}, we estimate the ECAL2/HCAL2 effect on the Primakoff trigger.
The mini- HCAL2 modules will have their gains well matched, and their analog signals
will be electronically summed and discriminated to provide a veto trigger signal for
hadron energies above 240 GeV.

\subsection{The Primakoff Trigger\label{sec:trigger}}

\begin{figure}[tbc]
\centerline{\rotate[r]{\epsfig{file=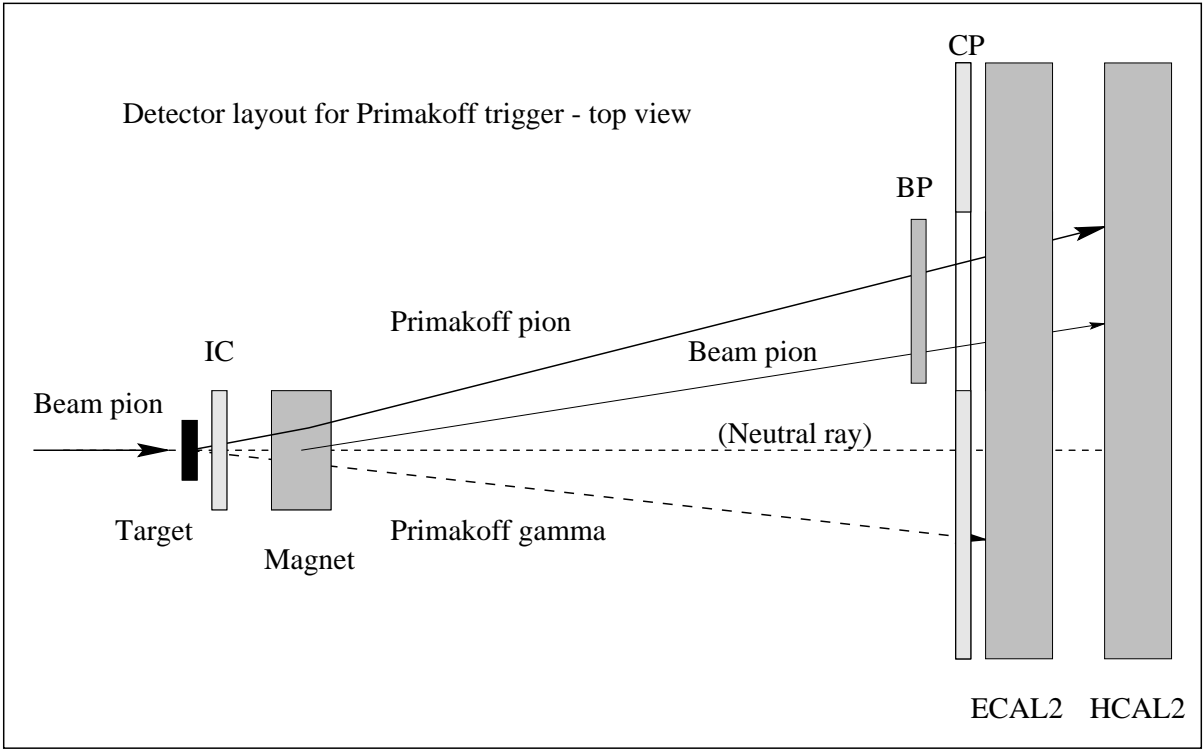,width=10cm,height=16cm}}}
\caption{Detector layout for Primakoff trigger.}
\label{fig:trigger}
\end{figure}

We construct the Primakoff trigger using two to three trigger levels. The 
main setup elements involved in the trigger are shown in Fig.~\ref{fig:trigger}. For
Primakoff physics, the trigger system electronics should ideally be localized in the
region of the ECAL2 and HCAL2 calorimeters, so that the final trigger signals may be
developed in the minimum time possible (to reduce dead time), and within the $300-1000$
nsec allowed in the DAQ.

The T0 trigger is a fast logical signal defining the beam phase space, rate and
purity at the target, and is generated near the target about 20 nsec after beam
passage. It is produced via a logical relation between signals from an ensemble of
beam transmission and beam halo veto (hole) scintillators located before the target.
Cuts on the analog signals of the transmission scintillators should reject upstream
interactions, beam spills with more than one particles, and thereby ensure a single
incident hadron at the target per event. The T0 trigger should ensure sufficient rate
with a small beam size and with a beam divergence below 1 mrad rms. The
size/divergence directly determines the performance of the T1 trigger downstream. For
example, a beam divergence of 1 mrad rms at target projected 40 m downstream (without
magnetic deflection) results in a beam spot size of 8 cm rms, and the stability of
the spot size and position is also important.

The T1 trigger exploits the essential feature of a Primakoff polarizability (and also
anomaly and hybrid) event: a coincidence between the $\gamma$ and $\pi$ detected in
ECAL2 and a scintillator BP respectively. Fig.~\ref{fig:trigger} shows the detectors
participating in this trigger. Proceeding downstream, we consider the scintillation
detectors IC (Interaction Counter), BP (Beam or Primakoff fiducial detector), CP
(Charged Particle), and the photon/hadron calorimeters ECAL2/HCAL2. The phototube
bases of all scintillators and IC counters should ensure gain stability at high
counting rate.

The IC counter logical signal should correspond to an amplitude of 1 MIP. Using two
successive IC counters allows an OR between their signals, which better accounts for
the Landau distribution. In addition, with two IC counters, consistency and stability
checks are possible.

BP is a scintillator fiducial paddle of dimensions 60 cm (in H) by 15 cm (in V), which
is the size covered by the beam and the Primakoff scattering events (pions or kaons).
BP is large enough in order to cover both the deflected beam and the scattered
Primakoff particles. It includes the beam region, for the reasons discussed in
section ~\ref{sec:how} above. BP helps form the pion detection trigger; it is set to
fire on a 1 MIP window condition. To accommodate also the anomaly and hybrid
triggers, the BP size was increased somewhat to account for the larger angular spread
of the $\pi^-$'s from these channels. Simulations in progress will help fix the
definitive sizes of BP, CP, and the ECAL2 hole.

CP is a charged particle veto scintillator array positioned at the front face of
ECAL2. It is designed with a hole slightly larger than the BP detector, in which the
BP detector above is positioned. It covers the front face of ECAL2. CP protects ECAL2
from charged leptons or hadrons directed at them. Before we use CP, we will make
measurements to assure that CP is unaffected by backsplashed charged particles from
ECAL2 and HCAL2. To reduce backsplash, one may possibly add some low Z material in
front of ECAL2. We require an ECAL2 energy deposit above 60 GeV and an HCAL2 energy
deposit below 240 GeV.

The first level trigger T1 is defined as:
\begin {equation}
T1 = IC (1~\rm{mip}) \cdot BP (1~\rm{mip}) \cdot \overline{CP}
         \cdot ECAL2 (> 60~\rm{GeV}) \cdot \overline{HCAL2} (>240~\rm{GeV}).
\end{equation}

The trigger is designed to accept only events in which one Primakoff scattered pion
hits and fires BP, the $\gamma$ energy exceeds 60 GeV, and HCAL2 measures less than
240 GeV. All of the non-interacting pion beam and most of the Primakoff scattered
pions pass through the ECAL2 hole (vertical size 7.6 cm, horizontal size 60 cm), and
then no charged particle hits ECAL2 anywhere. These pions proceed to HCAL2, where
their energy is measured well. Beam pions will then give HCAL2 signals greater than
240 GeV. Those Primakoff scattered pions that hit the blocks at the ECAL2 hole
periphery (the wall of fire, or WOF) will lose some energy in ECAL2, and will not
therefore have a good hadron energy measurement in HCAL2. We will reduce backgrounds
associated with the tail of the beam hitting ECAL2 blocks on the WOF, by omitting
these blocks from the ECAL2 sum signal. The ECAL2 low energy threshold is important
to suppress low energy backgrounds and the electronic noise of the analog sum signals.
In order to minimize electronic noise, both ECAL2 and HCAL2  summing  circuits should
use the techniques developed for this purpose of the GAMS experiment. This allows the
noise level to be significantly lower than that given by the sum of all channels.

The task of the first level trigger T1 is to provide a fast gate signal to start
digitization (for example in the ADC-system of the calorimeter) about $\sim$ 300 ns
after the beam traverses the target (see table~\ref{tab:rate}). The rate of this
signal (2.5~$10^3$ per spill in table ~\ref{tab:rate}) will be significantly lower
than the maximum of 1~$10^5$ per spill trigger rate accepted by the COMPASS data
acquisition.

A second level trigger T2 can be constructed if a faster T1 or more rate reduction is
needed. We will have a faster T1, if the IC counter 1 MIP trigger signal, which
arrives the latest at ECAL2, will be transferred from T1 to T2. Further rate
reduction may be gained using the additional trigger condition at T2 level. For
example, a momentum determination from the pion direction near the target and behind
SM2 can conceivably be integrated into the trigger scheme at a later stage, or at
least be used as a cross check of the HCAL2 energy trigger. The T2 trigger should
arrive no later than $\sim$ 1~ $\mu$sec after the beam traverses the target.

Further data reduction may be done in the COMPASS filter farm, before writing to
tape, using more detailed information from the tracking detectors and the ADCs.

\begin{table}
\begin{center}
\begin{tabular}[tbc]{|l|l|r|r|c|}
\hline
Signature & Range & Timing (nsec) & Reduc. Fact.& Rate (events/spill)\\
\hline
Beam                    & $-$           & 0   & $-$ & 1.25~10$^7$ \\
IC (interaction counter)& 1 mip         & 1   & $-$ & 1.25~10$^7$   \\
BP (beam or Primakoff)  & 1 mip         & 200 & $-$ & 1.25~10$^7$ \\
CP (charged particles)  & $\geq1$ mip   & 200 &   5 & 2.5~$10^6$ \\
HCAL2 ($\pi$ energy)    & $<240$ GeV  & 260 & $-$ & $-$   \\
                        &               &     & 1000  & 2.5~$10^3$     \\
ECAL2 ($\gamma$ energy) & $>60$ GeV  & 260 & $-$ & $-$   \\
\hline
\end{tabular}
\end{center}
\caption{The Primakoff trigger conditions and estimation of timing
relative to the target crossing time, and trigger rate reduction. For HCAL2
and ECAL2 we consider coincidences and a common reduction factor.}
\label{tab:rate}
\end{table}

\subsection{Expected Trigger Rates\label{sec:trig1}}

The ECAL2 $\gamma$ signal above 60 GeV in coincidence with BP and with HCAL2 $(40-240
\rm{GeV})$ should reduce the trigger rate from the beam rate by a factor of 1000. The
CP detector requirement should give at least another factor of 5 rate reduction. In
this way, one may expect to achieve a trigger rate lower than the beam rate by a
factor of 5000. The resulting rate is 20 times lower than the maximum of 10$^5$ per
spill DAQ limit in COMPASS.

We plan to study more precisely the background rates, and ways to reduce backgrounds.
For this purpose, we will use an event generator for pion-nucleus interactions,
embedded in the COMPASS apparatus. We will study what fraction of the events
generated pass our trigger conditions. The factor 1000 reduction above is only a
guess of what we expect from $\pi\gamma$ coincidence condition, with the energy
ranges of table~\ref{tab:rate}. The backgrounds will come from $\pi\gamma$
coincidences of non-Primakoff events, the rate of which depends on the ratio between
the total inelastic and photon production cross sections in the target.

\subsection{Measurement Significance}

The Primakoff Compton cross section is 1.2 mbarn for Pb, while the the total
inelastic cross section is 1.8 barn. For a  COMPASS pion beam rate of 5 MHz (or
1.25~10$^7$ particles/spill see table~\ref{tab:beam}), and a 1\% interaction Pb
target, we expect roughly 80 events/spill (80 $\approx 1.2/1.8 \times 10^{-3} \times
10^{-2} \times 1.25~10^{7}$) from the pion Primakoff effect. This corresponds to
10$^{7}$ events per month at 100\% efficiency. Assuming a trigger efficiency of 50\%
(due to the energy cuts), an accelerator operating efficiency of 50\%, and a tracking
efficiency of 80\%, one may expect to observe 2~10$^{6}$ Primakoff Compton events per
month. Statistics of this order will allow systematic studies, with fits carried out
for different regions of t, s, photon energy $\omega$, Z$^2$, etc.; and 
polarizability determinations with statistical uncertainties of order 0.2 .

For the kaon polarizability, due to the lower intensity, the statistics will be
roughly 50 times. A precision polarizability measurement requires more data taking
time. Comparing chiral anomaly to polarizability data, we expect roughly 300 times
statistics, due to the 140 times lower cross section and the lower $\pi^0$ efficiency
(\cite {ca}. The detailed simulations and count rates expected for these channels will
be presented in later reports.

\section{Budget}

We first consider the needs of the ECAL2 and HCAL2 calorimeters. To obtain good
acceptance for all the Primakoff reactions discussed above, an ECAL2 with size of
order 1.5 m diameter is required. The optimum size is under study. This may be
achieved with GAMS-4000 lead glass blocks, each of size $3.8\times3.8$ cm$^2$. An
HCAL2 calorimeter with 100 20$\times$20 cm$^2$ Dubna cells can satisfy beam killer
and PID functions. The ADC's for these blocks must be designed and built.

\section{Conclusions}

A hadron Primakoff physics program for COMPASS is proposed. The physics topics $-$
measurement of the pion polarizability, search for hybrid meson(s), studies of the
chiral anomaly and of radiative meson transitions $-$ are discussed with emphasis on
the pion polarizability, for which the simulations and detector studies are most
advanced. The program can be achieved in a COMPASS run, using negative and positive
hadron beams.

The proposed program can be run with a  partially instrumented COMPASS spectrometer,
consisting of the spectrometer magnets, the central tracking detectors, (parts of)
the ECAL2/HCAL2 calorimeters and a relatively simple trigger. All physics topics
proposed can be studied simultaneously.

\section{Acknowledgments}

This research was supported by the U.S.-Israel Binational Science Foundation research
was supported by the U.S.-Israel Binational Science Foundation and the Israel Science
Foundation founded by the Israel Academy Sciences and Humanities, Jerusalem, Israel.
The authors thank M. Finger and M. Chavleishvili for the friendly atmosphere at the
Charles U./JINR (Dubna)/International U. (Dubna) Prague COMPASS 1997 summer school.
Thanks are due to M.P.I. Heidelberg SELEX/COMPASS group, U. Dersch, F. Dropmann, I.
Eschrich, Kruger, J. Pochodzalla, B. Povh, J. Simon, and K. Vorwalter, for hospitality
collaboration during the writing of this report. Thanks are due M. Buenerd, P.
Cooper, D. Drechsel, T. Ferbel, L. Frankfurt, A. Ocheraschvili, S. Paul, J. Russ, I.
Savin, H.W. Siebert, A. Singovsky, N. Terentyev, U. Wiedner, and T. Walcher for
valuable  discussions.




\end{document}